\documentclass{article}
\usepackage[final,nonatbib]{neurips_2023}

\usepackage[utf8]{inputenc} % allow utf-8 input
\usepackage[T1]{fontenc}    % use 8-bit T1 fonts
\usepackage[hidelinks]{hyperref}       % hyperlinks
\usepackage{url}            % simple URL typesetting
\usepackage{booktabs}       % professional-quality tables
\usepackage{amsfonts}       % blackboard math symbols
\usepackage{nicefrac}       % compact symbols for 1/2, etc.
\usepackage{microtype}      % microtypography
\usepackage{xcolor}         % colors
\usepackage{amsmath}
\usepackage{amssymb}
\usepackage{bbm}
\usepackage{mathtools}
\usepackage{soul}
\usepackage{cleveref}
\usepackage[labelfont=bf]{caption}
\usepackage{appendix}
\usepackage{multicol}

\usepackage[
    style=numeric-comp,
    sorting=none,
    doi=false,
    url=false,
    eprint=false,
    maxbibnames=21,
]{biblatex}
\AtEveryBibitem{\clearlist{language}}

\def\bs{\boldsymbol}
\def\R{\mathbb{R}}
\def\E{\mathbb{E}}
\def\N{\mathcal{N}}

\DeclareRobustCommand{\mb}[1]{\boldsymbol{#1}}

\renewcommand{\mid}{~\vert~}

\newcommand{\mbm}{\mb{m}}

\newcommand{\mbx}{\mb{x}}
\newcommand{\mby}{\mb{y}}

\newcommand{\mbI}{\mb{I}}

\newcommand{\mbL}{\mb{L}}

\newcommand{\mbU}{\mb{U}}

\newcommand{\mbmu}{\mb{\mu}}

\newcommand{\mbLambda}{\mb{\Lambda}}

\newcommand{\mbSigma}{\mb{\Sigma}}

\newcommand{\cG}{\mathcal{G}}

\newcommand{\cN}{\mathcal{N}}

\newcommand{\cP}{\mathcal{P}}

\newcommand{\cW}{\mathcal{W}}
\newcommand{\cX}{\mathcal{X}}

\newcommand{\reals}{\mathbb{R}}

\title{Estimating Noise Correlations Across Continuous Conditions
With Wishart Processes}

\author{%
  Amin Nejatbakhsh \qquad Isabel Garon \qquad Alex H Williams\\
  Center for Neural Science, New York University, New York, NY\\
  Center for Computational Neuroscience, Flatiron Institute, New York, NY\\
  \texttt{\{anejatbakhsh,igaron,awilliams\}@flatironinstitute.org} 
}
\begin{document}

\maketitle

\begin{abstract}
% The signaling capacity of a neural population depends on the scale and orientation of its covariance across trials.
% Estimating this ``noise'' covariance is challenging and is thought to require a large number of stereotyped trials.
% New approaches are therefore needed to interrogate the structure of neural noise across rich, naturalistic behaviors and sensory experiences, with few trials per condition.
% Here, we exploit the fact that conditions are smoothly parameterized in many experiments, and leverage Wishart process models to pool statistical power from trials in neighboring conditions.
% We demonstrate that these models performs favorably on experimental data from mouse visual cortex and monkey motor cortex relative to standard covariance estimators.
% Moreover, they produce smooth estimates of covariance as a function of stimulus parameters, enabling estimates of noise correlations in entirely unseen conditions as well as continuous estimates of Fisher information---a commonly used measure of signal fidelity.
% Together, our results suggest that Wishart processes are a flexible, broadly applicable, and feasible approach to quantify trial-to-trial neural variability in a Bayesian framework.

The signaling capacity of a neural population depends on the scale and orientation of its covariance across trials. Estimating this ``noise'' covariance is challenging and is thought to require a large number of stereotyped trials. New approaches are therefore needed to interrogate the structure of neural noise across rich, naturalistic behaviors and sensory experiences, with few trials per condition. Here, we exploit the fact that conditions are smoothly parameterized in many experiments and leverage Wishart process models to pool statistical power from trials in neighboring conditions. We demonstrate that these models perform favorably on experimental data from the mouse visual cortex and monkey motor cortex relative to standard covariance estimators. Moreover, they produce smooth estimates of covariance as a function of stimulus parameters, enabling estimates of noise correlations in entirely unseen conditions as well as continuous estimates of Fisher information---a commonly used measure of signal fidelity. Together, our results suggest that Wishart processes are broadly applicable tools for quantification and uncertainty estimation of noise correlations in trial-limited regimes, paving the way toward understanding the role of noise in complex neural computations and behavior.
\end{abstract}

\section{Introduction}
\label{sec:intro}

\begin{figure}
\includegraphics[width=\linewidth]{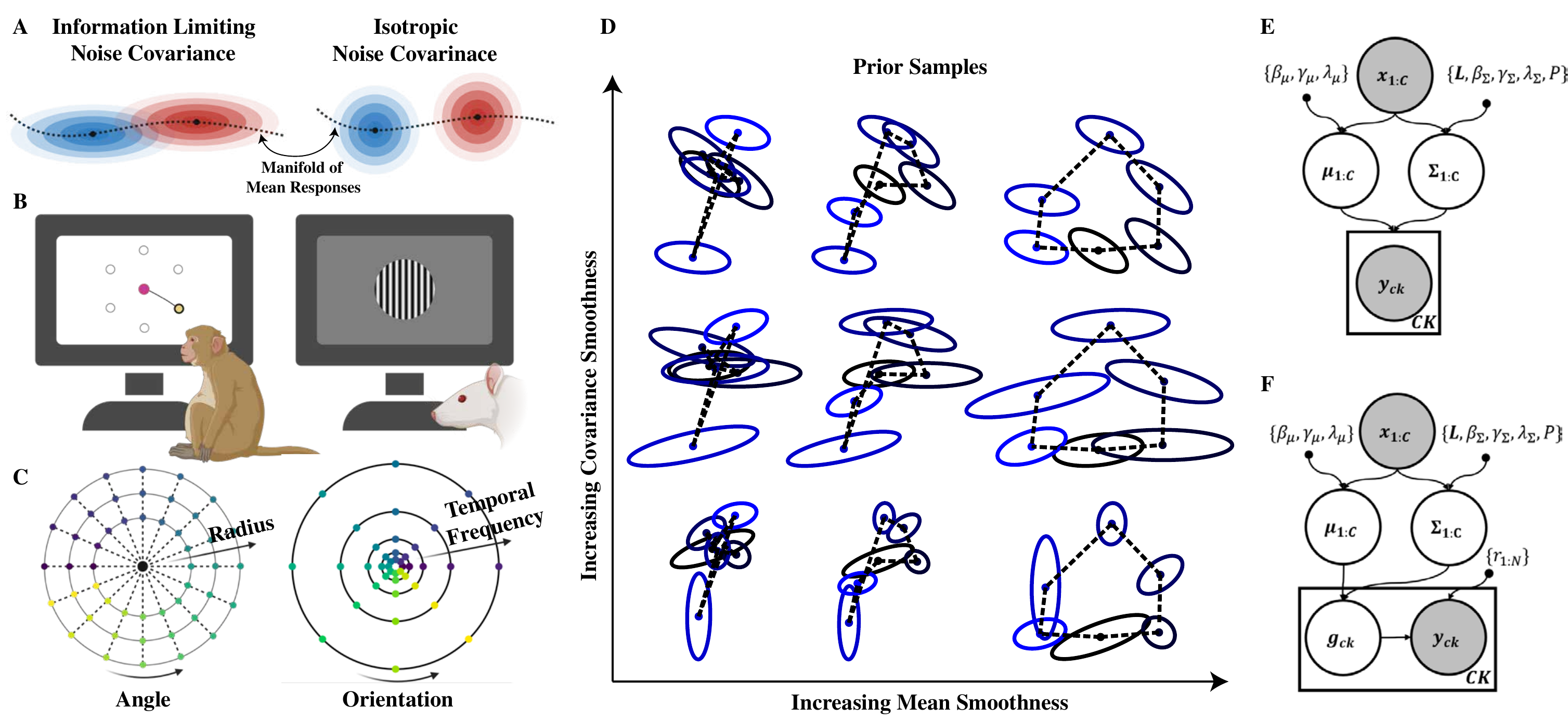} 
\caption{(A) Illustration of information limiting noise correlations.
(B) Experimental datasets with smoothly parametrized conditions~(see \cite{Moreno2014}).
\textit{Left}, a nonhuman primate makes point-to-point reaches to radial targets.
\textit{Right}, a rodent views drifting visual grating stimuli of different orientations and temporal frequencies (i.e. speed).
(C) Parameterized condition spaces for the datasets in B.
(D) Samples from Gaussian and Wishart process prior distributions for with $C=6$ conditions over 1D periodic variable (e.g. angle) and $N=2$ neurons at varying kernel smoothness parameters ($\lambda$ in eq.~\ref{eq:kernel-periodic}).
Dots and ellipses represent the mean and covariance of neural responses.
Colors appear in the increasing order of condition value (e.g. angle) and dashed lines connect neighboring conditions. Increasing $\lambda_{\mu}$ 
% in $k_\mu$
(horizontal axis) encourages smoothness in the means while increasing $\lambda_{\Sigma}$ 
% in $k_\Sigma$
(vertical axis) encourages the ellipses to change smoothly in scale and orientation.
(E-F) Graphical model of the Wishart model with Normal (E) and Poisson (F) observations (see Supplement~\ref{supp:poisson} for details of the Poisson model).
}
% \vspace{-1em}

\label{fig:schematics}
\end{figure}

Nominally identical repetitions of a sensory stimulus or a behavioral action often elicit variable responses in neurons.
Intuitively, this ``noise'' degrades the information capacity of neural representations.
However, the precise nature of this effect is complex since neural responses are not independent random events.
Indeed, pairs of co-recorded neurons exhibit significant ``noise correlations'' from trial-to-trial, with well-studied and intriguing consequences~\cite{Zohary1994-ns,Shadlen1996-dr,averbeck2006neural,Cohen2011-ke,Panzeri2022}.
In particular, introducing noise correlations into a circuit can theoretically degrade or enhance its signaling capacity depending on the geometric configuration of neural representations in firing rate space
(Fig.~\ref{fig:schematics}A; \cite{Abbott1999-tb,Moreno2014}).

Estimating noise covariance is therefore an important problem that arises frequently in neural data analysis.\footnote{Throughout, we focus on the problem of estimating noise \textit{covariance} instead of noise \textit{correlation}. By accurately estimating the former, we can also estimate the latter since the correlation between two neurons is given by their covariance normalized by each neuron's marginal variance.}
However, this problem is widely regarded as challenging~\cite{Yatsenko2015-vy,Williams2021-single-trial-limit,Rumyantsev2020-ar} because the number of estimated parameters (i.e., number of unique neuron pairs) grows quadratically with the number of co-recorded neurons, while the number of measurements per trial (i.e., number of neurons with a response) grows only linearly (for more discussion see, e.g., \cite{Johnstone2018}).
As a result, experimental investigations into noise correlation have been limited to extremely simple tasks where a very large number of trials can be collected over a small number of conditions.
For example, \textcite{Rumyantsev2020-ar} recorded neural responses over just two stimulus conditions (oriented visual gratings), repeated across hundreds of trials.
% We therefore have a rudimentary characterization of how noise correlations change across rich or naturalistic experimental conditions, despite experimental evidence that noise statistics are context-dependent~\cite{Kohn2005,Cohen2008,Alvarez2013} and interest among the theoretical community~\cite{Dapello2021,Panzeri2022,Duong2023}.
We therefore have a rudimentary characterization of how noise correlations change across rich or naturalistic experimental conditions, despite interest among the theoretical community~\cite{Dapello2021,Panzeri2022,Duong2023}, and experimental evidence that noise statistics are context-dependent~\cite{Kohn2005,Cohen2008,Alvarez2013}.
New statistical approaches that are capable of inferring noise covariance with fewer trials per condition are needed to make progress on this subject.

To address this need, we develop a probabilistic model based on Wishart processes~\cite{Gourieroux2009-gp}.
The core insight we exploit is that similar experimental conditions---e.g. visual gratings at similar orientations or cued arm reaches to similar locations---ought to exhibit similar noise statistics.
Thus, by building this insight into the prior distribution of a probabilistic model, we can pool statistical power across neighboring conditions and accurately infer noise correlation structure with fewer trials per condition.
In fact, this principle also enables us to predict covariance in entirely unseen conditions by interpolating between (and potentially extrapolating from) observed conditions.
We demonstrate the scientific potential and generality of this approach through two experimental applications spanning different modalities (vision and motor tasks) and species (mouse and nonhuman primate).

\section{Methods}

\subsection{Problem Setup}
\label{subsec:related-work}

Consider an experiment with $N$ neurons recorded over $C$ distinct stimulus conditions, each with $K$ repeated trials.\footnote{To simplify exposition, we assume that there are an equal number of trials in each condition. However, the model easily applies to the more general case with unbalanced trial allocations.}
Experimental advances continually enable us to record larger neural populations (large $N$) and neuroscientists are increasingly interested in performing rich and naturalistic experiments (large $C$).
Both of these trends put us in a regime where $K < N$, meaning there are few trials per condition relative to the number of neurons~\cite{Williams2021-single-trial-limit}.

\paragraph{Empirical Estimator:} Our goal is to estimate the $N \times N$ covariance matrix, $\mbSigma_c$, describing trial-by-trial variability in each condition ${c \in \{1, \hdots, C\}}$.
If $\mby_{ck} \in \reals^N$ is the measured response on trial $k$ of condition $c$, then a straightforward estimator is the empirical covariance:
\begin{equation}
\label{eq:empirical-covariance}
\overline{\mbSigma}_c = (1/K)\sum_k (\mby_{ck} - \overline{\mby}_c) (\mby_{ck} - \overline{\mby}_c)^\top~,
\quad \text{where} \quad
\overline{\mby}_c = (1/K) \sum_k \mby_{ck}~.
\end{equation}
This estimator behaves poorly in our regime of interest when $K < N$.
Indeed, while the true covariance must be full rank, the empirical covariance is singular with $N - K$ zero eigenvalues.

\paragraph{Regularized Estimators:} Statisticians often refer to problems where $K < N$ as the \textit{high-dimensional regime}.
Estimating covariance in this regime is actively studied and understood to be challenging \cite{Pourahmadi2013}.
The Ledoit-Wolf estimator \cite{Ledoit2004} is a popular approach, which regularizes (``shrinks'') the empirical covariance towards a diagonal estimate.
Other estimators shrink the individual covariance estimates towards the grand covariance (see eq.~\ref{eq:grand-empirical-covariance-factorization} below) instead of a diagonal target~\cite{rahim2019population}.
% Other estimators apply regularization to the eigenvalue spectrum of $\overline{\mbSigma}_c$ with a similar goal of shrinking the empirical spectrum towards a uniform distribution~\cite{Donoho2018}.
Graphical LASSO~\cite{Friedman2007} is another popular approach, which incorporates sparsity-promoting penalty on the inverse covariance matrix.
% 
% \amin{
% Closest to our framework is a recent method called PoSCE which posits a probabilistic model over the tangent space of the PSD manifold combining the per-condition empirical covariances with the grand covariance across conditions
% and uses a geometric mean to perform the averaging~\cite{rahim2019population}.
% }
% 
All of these methods can be understood as forms of regularization, which reduce the variance of the covariance estimate at the expense of introducing bias.

\paragraph{Grand Empirical Estimator:} The methods summarized above (or variants thereof) are common within the neuroscience literature~\cite{Yatsenko2015-vy,Cai2019-pi,Li2021}.
It is also common to simply estimate the covariance as being constant across conditions~\cite{Rumyantsev2020-ar,Walther2016-se}, in which case an estimate can be found by pooling trials across all conditions to achieve an empirical estimate of the grand covariance:
\begin{equation}
\label{eq:grand-empirical-covariance-factorization}
\overline{\overline{\mbSigma}} = \frac{1}{C} \sum_c \overline{\mbSigma}_c = \frac{1}{KC} \sum_{k,c} (\mby_{ck} - \overline{\mby}_c) (\mby_{ck} - \overline{\mby}_c)^\top.
\end{equation}
This estimator can be thought of as having high bias (since covariance is often thought to be stimulus-dependent~\cite{Kohn2005,Cohen2008,Alvarez2013}), but low variance since many more trials can be averaged over after pooling across conditions.
We compare our model to these baselines in section~\ref{sec:results}.

\subsection{Wishart Process Model of Noise Correlations}
\label{subsec:wishart-process-model-methods}

Our insight is to incorporate an alternative form of regularization whenever the experimental conditions are smoothly parameterized.
For instance, a classical stimulus set used by visual neuroscientists is a grating pattern varied over $C$ orientations.
In fact, such stimuli are commonly used for noise correlation estimation~\cite{Yatsenko2015-vy,Rumyantsev2020-ar,Stringer2021,Kafashan2021}.
Intuitively, two gratings with nearly identical orientations should evoke similar probabilistic responses.
Thus, as stimulus orientation is incremented, the mean and covariance of the response should smoothly change along a manifold (Fig.~\ref{fig:schematics}A).

This idea readily extends to other experimental settings.
For example, pure acoustic tones with smoothly varied frequencies are often used in auditory neuroscience~\cite{Merzenich1975-ko}.
In studies of olfactory coding, isolated odorants or mixtures of odorants can be presented at smoothly varied concentrations~\cite{Kudryavitskaya2021-ky}.
In motor neuroscience, voluntary reach movements can be cued to continuously varied target locations~\cite{Even-Chen2019-fi}.
Finally, in the context of navigation, noise correlations have also been studied in neural ensembles coding for continuous variables like position~\cite{Hazon2022-ff} and head direction~\cite{Peyrache2015-ov}.
In principle, the modeling approach we develop is applicable to any of these experimental settings. This is in contrast to experimental settings that employ a discrete set of stimuli with no underlying continuous parameterization (such as olfactory responses to distinct classes of odors).

To formalize these concepts, let $c \in \{1, \dots, C\}$ index the set of conditions, and let $\mbx_c \in \cX$ denote the stimulus or animal behavior in condition $c$.
For example, $\mbx_c$ could be a scalar denoting the orientation of a voluntary reach or a visual grating~(Fig.~\ref{fig:schematics}B).
However, the model we will describe is more general than this: $\mbx_c$ could be a vector describing a multi-dimensional stimulus space.
For example, we will later study motor preparatory activity to reach targets at different radii and angles (sec.~\ref{subsec:reaching-results}).
In this setting, $\mbx_c$ is a 2-dimensional vector specifying these two parameters of the reach target (Fig.~\ref{fig:schematics}C, left).
We will also study neural responses to drifting visual gratings (sec.~\ref{subsec:visual-gratings-results}).
In this setting, $\mbx_c$ is a vector specifying the speed and orientation of the grating stimulus (Fig.~\ref{fig:schematics}C, right).

Recall that $\mby_{ck} \in \reals^N$ denotes the simultaneously measured response of $N$ neurons to condition $c \in \{1, \dots, C\}$ on trial $k \in \{1, \dots, K\}$.
Our goal is to model these measured responses as samples from some distribution $F$ with mean $\mbmu$ and covariance $\mbSigma$:
\begin{equation}
\mby_{ck} \sim F(\mbmu(\mbx_c), \mbSigma(\mbx_c)) \quad \text{independently for $c, k \in \{1, \dots C \} \times \{1, \dots K\}$.}
\label{eq:observation-model}
\end{equation}
Importantly, the functions $\mbmu : \cX \mapsto \reals^N$ and $\mbSigma : \cX \mapsto \reals^{N \times N}_{++}$ should be smooth so that the response distribution changes gradually as the experimental condition is altered.
To accomplish this, we leverage the machinery of Gaussian processes, which can be used to specify a prior distribution over smooth functions~\cite{Williams2006-gps}.
While Gaussian process models have been extensively used within statistical neuroscience~\cite{Yu2008,Duncker2019-zm,keeley2020efficient,Rutten2020,Keeley2020identifying}, they have typically been used to model smoothness in temporal dynamics rather than smoothness across experimental conditions (but see \cite{Wu2017}).

In this paper, we assume that $F(\mbmu, \mbSigma)$ is a multivariate Gaussian distribution, which enables tractable and convenient probabilistic inference (see sec.~\ref{subsection:methods-variational-inference}).
Since many noise correlation analyses are applied to spike count data, it is of great interest to extend our model so that $F$ is better suited to this case.
Multivariate extensions of the Poisson distribution are an attractive option \cite{Inouye2017}.
While the focus of this paper is on multivariate Gaussian distributions, we develop inference tools for the Poisson model and we leave further investigation of this model for future work (see~\ref{fig:schematics}F for the corresponding graphical model and Supplement~\ref{supp:poisson} for further details).
% However, such extensions are non-trivial and outside the scope of our initial investigation into this model class.
We note that it is common to pre-process spike count data and calcium fluorescence traces (e.g. through square root or log transforms) so that they are better approximated by Gaussian models~\cite{Yu2008,Buzsaki2014-ea,Zhu2022-ey}.

To model $\mbmu(\cdot)$, we stack independent Gaussian processes into an $N$-vector.
To model $\mbSigma(\cdot)$, we use a \textit{Wishart process}~\cite{Wilson2011}.
In essence, the Wishart process (denoted by $\mathcal{WP}$) stacks Gaussian processes into an $N \times P$ matrix and uses this as an overcomplete basis for the covariance. Formally,
\begin{equation}
\mbmu(\cdot) \sim \cG\cP^N(\bs{0}, k_\mu) ~, \quad
\mbU(\cdot) \sim \cG\cP^{N \times P}(\mb{0}, k_\Sigma) ~, \quad
\mbSigma(\mbx) = \mbU(\mbx) \mbU(\mbx)^\top,
\label{eq:gaussian-wishart-priors}
\end{equation}
where $P \geq N$ is a hyperparameter, and $\cG\cP^N(\mbm, k)$ denotes an $N$-dimensional Gaussian process with (constant) mean $\mbm \in \reals^N$ and a positive definite kernel function ${k : \cX \times \cX \mapsto \reals}$, described below.
For any fixed value of $\mbx \in \cX$, the product $\mbU(\mbx) \mbU(\mbx)^\top$ follows a Wishart distribution, which is a standard probabilistic model for covariance and inverse covariance~(see sec. 3.6 of \cite{gelman2013bayesian}).
To our knowledge, Wishart processes have not previously been leveraged for neural data analysis.

The kernel functions $k_\mu$ and $k_\Sigma$ play important roles by respectively specifying the degree of smoothness in the mean and covariance of the neural response across conditions (Fig.~\ref{fig:schematics}D).
The kernel functions must satisfy certain technical conditions, as reviewed in~\cite{Scholkopf2001,Hofmann2008}.
We satisfy these conditions by defining a kernel for each coordinate of $\mbx \in \cX$ and evaluating the product, $k(\mbx, \mbx^\prime) = \prod_i k_i(x_i, x^\prime_i)$.
This is guaranteed to produce a valid kernel (\cite{Scholkopf2001}, Proposition 13.2).
For non-periodic variables, we use a standard squared exponential kernel
\begin{equation}
\label{eq:kernel-squared-exp}
k(x, x^\prime) = \gamma \delta_{x x^\prime} + \beta \exp \left [ -(x - x^\prime)^2  / \lambda \right ]
\end{equation}
where $\{\gamma, \beta, \lambda \}$ are hyperparameters and $\delta_{xx^\prime} = 1$ when $x = x^\prime$ and otherwise evaluates to zero.
For periodic conditions (e.g. orientation of visual gratings or radial reach directions) we use~\cite{Mackay1998}:
\begin{equation}
\label{eq:kernel-periodic}
k(x, x^\prime) = \gamma \delta_{x x^\prime} + \beta \exp \left [ -\sin^2 (\pi |x - x^\prime| / T)/ \lambda \right ]    
\end{equation}
where $T$ specifies the period of the kernel.
For both kernels, the most critical hyperparameter is $\lambda$ (often called the ``bandwidth''), which specifies the degree of smoothness across conditions.
The remaining hyperparameters, $\gamma$ and $\beta$, respectively ensure that the Gaussian process is well-conditioned and scaled.
We tune all of these hyperparameters to maximize heldout log-likelihood by cross-validation.
Throughout, we use $\{\gamma_\mu, \beta_\mu, \lambda_\mu\}$ and $\{\gamma_\Sigma, \beta_\Sigma, \lambda_\Sigma\}$ to respectively represent the kernel hyperparameters for $k_\mu$ and $k_\Sigma$.

\subsection{Low-Rank Plus Diagonal Covariance and Other Model Variants}

The vanilla Wishart process model in equations (\ref{eq:observation-model}) and (\ref{eq:gaussian-wishart-priors}) can be extended in several ways, which we found to be useful in practical applications. 
First, we can augment the covariance model by adding a diagonal term, resulting in a model that is reminiscent of factor analysis:
\begin{equation}
\label{eq:wishart-with-diag-component}
\mbSigma(\mbx) = \mbU(\mbx) \mbU(\mbx)^\top + \mbLambda(\mbx),
\end{equation}
where $\mbLambda(\mbx) \in \reals^{N \times N}$ is a condition-dependent nonnegative diagonal matrix where the diagonal terms are independent samples from a $\cG\cP^N$ that are transformed using a \texttt{softplus} function, ${f(x) = \log(1 + e^x)}$, to ensure positivity.
We term this extension of $\mathcal{WP}$ as $\mathcal{WP}^{lrd}$ (for low-rank plus diagonal).
Intuitively, this model extension has fewer parameters than a vanilla $\cW\cP$ and hence discourages overfitting.
Additionally, the diagonal term, $\mbLambda(\mbx)$, helps ensure that the covariance matrices remain well-conditioned during inference.
% This model has several potential benefits: (a) the learnable diagonal term, $\Lambda(\mbx)$, ensures that matrices are well-conditioned during inference, (b) it has fewer parameters than the vanilla $\cW\cP$ and hence discourages overfitting.
% 
% In practice, this diagonal term keeps the covariance matrix well-conditioned during latent variable inference (see section \ref{subsection:methods-variational-inference}).
% Even though covariance should be invertible in theory whenever $P \geq N$, we found that setting $P$ equal to or slightly above $N$, the inferred covariance often becomes ill-conditioned during optimization.
Indeed, we can choose values of $P < N$ (or even $P=0$), which enables us to specify prior distributions over the covariance that are ``low-dimensional'' in the sense that a large fraction of variance can be explained by a small number of dimensions.
This comports with numerous experimental observations~\cite{Rabinowitz2015,Rumyantsev2020-ar}.
As can be seen in Supplement Fig.~\ref{suppfig:low-rank-supplement}, the dimensionality of matrices sampled from a vanilla Wishart distribution scales linearly with $N$, while matrices sampled according to equation (\ref{eq:wishart-with-diag-component}) saturate at lower values.

Next, we additionally incorporate a lower-triangular matrix $\mbL \in \reals^{N \times N}$ into the covariance model to capture aspects of covariance that are condition-independent (i.e. do not depend on $\mbx$):
\begin{equation}
\label{eq:wishart-with-mean-and-diag-component}
\mbSigma(\mbx) = \mbL (\mbU(\mbx) \mbU(\mbx)^\top + \mbLambda(\mbx)) \mbL^\top,
\end{equation}
In the literature $\mbL\mbL^T$ is referred to as the scale matrix.
Some intuition can be gained by assuming $\mbLambda(\mbx) = \mb{0}$ and $P > N$, in which case $\mbU(\mbx) \mbU(\mbx)^\top \propto \mbI$ in expectation under the marginal prior distribution.
Thus, a good choice is to set $\mbL$ to be the Cholesky factor of the grand empirical covariance as defined in eq.~(\ref{eq:grand-empirical-covariance-factorization}).
In our experiments, we use this as an initialization for $\mbL$ and we optimize it to maximize the evidence lower bound (ELBO) as described in the Supplement~\ref{supp:inference}.

It is also possible, and potentially beneficial, to instead model smoothness in the inverse covariance matrix (also known as the precision matrix).
This results in an \textit{inverse Wishart process}.
% While the results in this paper are based on the Wishart process, the code package accompanying this paper supports both Wishart and inverse Wishart models. 
We refer the reader to Supplement~\ref{supp:inverse_wishart} for additional details of the inverse Wishart model.

\subsection{Latent Variable and Parameter Inference}
\label{subsection:methods-variational-inference}

Each of the probabilistic models above specifies a posterior distribution of the noise covariance conditioned on observed neural activity, $p \left ( \big \{ \mbSigma(\mbx_c) \big \}_{c=1}^C \mid \{\mby_{ck}\}_{c=1,k=1}^{C, K} \right )$.
More precisely, each model specifies a posterior on the latent covariance factor matrices, $p \left ( \big \{ \mbU(\mbx_c)  \big \}_{c=1}^C \mid \{\mby_{ck}\}_{c=1,k=1}^{C, K} \right )$, which defines the desired posterior over covariance through a measurable transformation (eq.~\ref{eq:wishart-with-mean-and-diag-component}).
This distribution over latent covariance factors has $NPC$ dimensions, precluding us from representing the posterior exactly (doing so would require us to tile $\reals^{NPC}$ with a lattice of points).
We instead use mean field variational inference~\cite{Blei2017-mm} to approximate the desired posterior.
Since this approach has already been described for similar Wishart process models~\cite{Heaukulani2019}, we relegate a description to Supplement~\ref{supp:inference}.
All codes are implemented in \texttt{numpyro}~\cite{numpyro} and available at \url{https://github.com/neurostatslab/wishart-process}.

\subsection{Connections to Population Decoding}
\label{subsection:methods-decoding}

As we discussed in \cref{sec:intro}, the estimation of noise covariance matrices is an important step for a variety of downstream neural analyses.
Here, we briefly highlight two applications of Wishart process models in the context of neural decoding and perceptual discrimination~\cite{averbeck2006neural}.

\paragraph{Quadratic Discriminant Analysis (QDA)}
A simple decoding task is to infer which condition $c \in \{1, \dots, C\}$ gave rise to an observed neural response, $\mby \in \reals^N$, on a single trial.
A natural approach is to use a maximum likelihood classifier, $\label{eq:qda}\texttt{class}(\bs{y}) = \arg \max_{c\in\{1, \dots, C\}} \log p(\bs{y}|\hat{\mbmu}_c,\hat{\mbSigma}_c)$.
Doing so requires that we construct estimates for the mean responses, $\hat{\mbmu}_1, \dots, \hat{\mbmu}_C$, and covariances, $\hat{\mbSigma}_1, \dots, \hat{\mbSigma}_C$.
Since covariance estimation is challenging, a common practice (see e.g.~\cite{Rumyantsev2020-ar}) is to use the grand empirical covariance (eq.~\ref{eq:grand-empirical-covariance-factorization}) and therefore set $\hat{\mbSigma}_1 = \dots = \hat{\mbSigma}_C$.
Under this model, known as \textit{linear discriminant analysis} (LDA; see e.g.~\cite{murphy2012machine}~Ch. 4), it can be shown that the decision boundaries of the maximum likelihood classifier become linear.

Intuitively, if we can accurately estimate distinct covariances for each condition, we may unlock more powerful decoders by relaxing the assumption of equal covariance across conditions.
Under this model, known as \textit{quadratic discriminant analysis} (QDA), the decision boundaries are described by hyperbolas (A schematic is shown in Fig.~\ref{fig:synthetic-data}D).
Furthermore, \textcite{pagan2016neural} described a circuit implementation of QDA using a cascade of two linear-nonlinear models, suggesting that such classifiers are a simple and biologically plausible neural computation.
We demonstrate the ability of Wishart process models to enable accurate decoding via QDA in Fig.~\ref{fig:synthetic-data}E.

% For each test trial, we assigned to it the class label that generated the maximum probability for that trial according to a QDA model with covariances estimated by different methods: $\label{eq:qda}\texttt{class}(\bs{y}) = \underset{c}{\arg \max} \log p(\bs{y}|\bs{\mu}_c,\bs{\Sigma}_c)$. The schematic of the classification and the QDA results are presented in Fig.~\ref{fig:synthetic-data}E,F.
% 
\paragraph{Fisher Information (FI)} A quantity that plays an important role in our understanding of how neural noise properties affect behavioral discriminability and representation is FI defined as $\E  [\partial^2 \log p(\bs{x}) / \partial \bs{x}^2]$ ~\cite{averbeck2006neural}.
% 
% 
% Unfortunately, due to the lack of a robust estimator most of the applications of FI have been limited to theoretical research. 
% 
Unfortunately, estimating FI is challenging because it involves measuring infinitesimal changes in the neural response distribution with respect to the stimulus condition (Fig.~\ref{fig:synthetic-data}F).
Wishart processes enable us to exploit the continuous parametrization of experimental conditions to derive a natural estimator for FI.
For constant covariances across conditions, the FI for multivariate normal distributions is given by $\bs{\mu}'(\bs{x})^T \bs{\Sigma}^{-1} \bs{\mu}'(\bs{x})$ where $\bs{\mu}'(\bs{x})$ is the gradient of the mean with respect to $\mbx$.
If covariances vary across conditions, this quantity is replaced with $\bs{\mu}'(\bs{x})^T \bs{\Sigma}^{-1}(\bs{x}) \bs{\mu}'(\bs{x}) + \frac{1}{2}\texttt{tr}([\bs{\Sigma}^{-1}(\bs{x}) \bs{\Sigma}'(\bs{x})]^2)$ where $\bs{\Sigma}'(\bs{x})$ is the gradient of the covariance w.r.t. the conditions and $\texttt{tr}$ is the trace operator.
Wishart process models enable the continuous estimation of FI as a function of $\mbx$. This is achieved by sampling from the gradient of the inferred posterior distribution (see Supplement~\ref{supp:fi}).

\section{Results}
\label{sec:results}
\begin{figure}
\includegraphics[width=\linewidth]{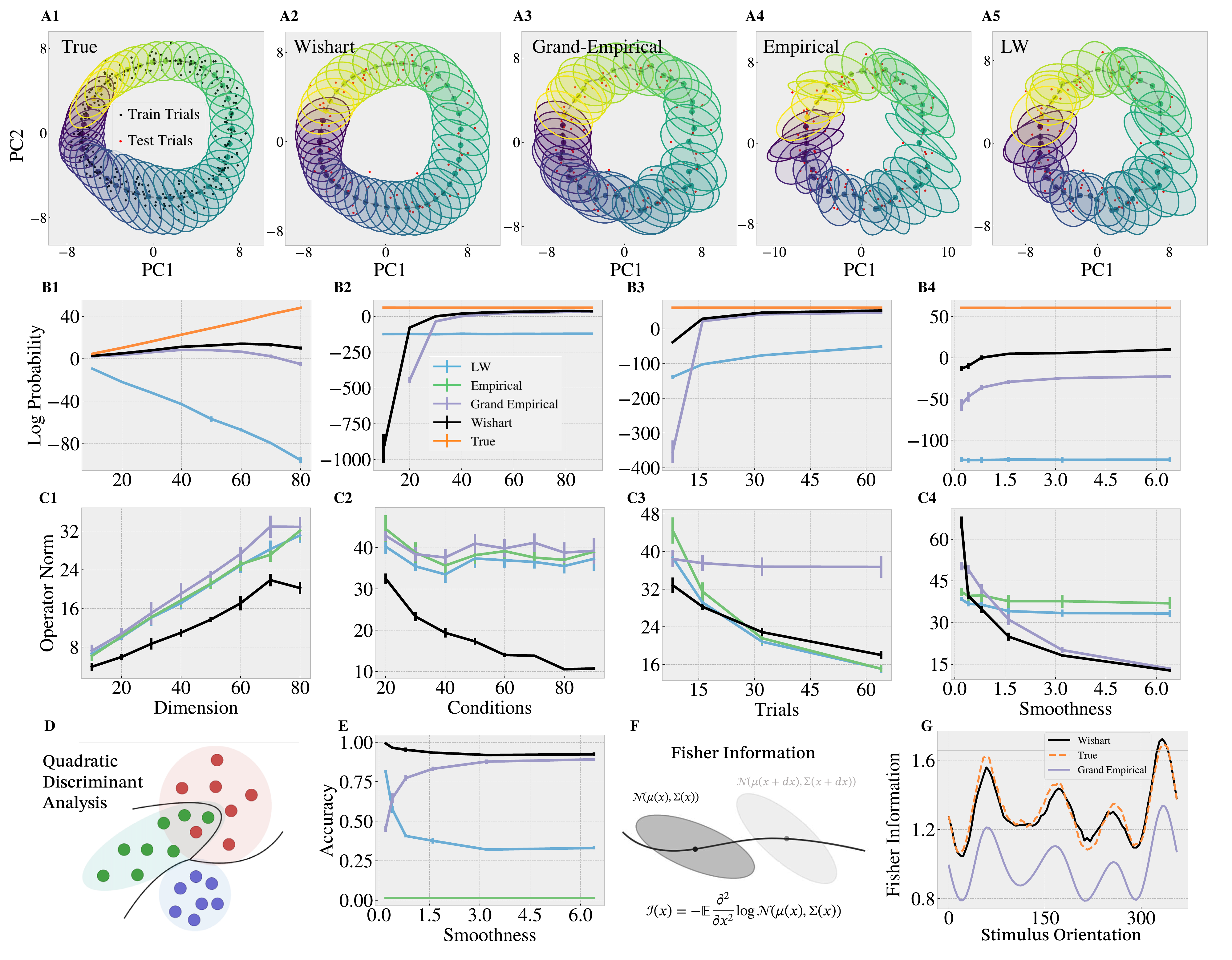} 
\caption{
Results on synthetic data. See Supplement~\ref{supp:simulation_details} for all simulation details.
(A) Simulated (A1) and inferred (A2-A5)
response distributions using different methods.
Means $\mbmu(\mbx_{c})$ and covariances $\mbSigma(\mbx_{c})$ are shown in dashed lines and colored ellipses in PC space. Train and test trials $\mby_{ck}$ are shown in black and red dots.
% 
% (B) Covariances inferred by different methods for two example conditions.
% 
(B1-B4) Mean log-likelihood of held-out trials as a function of the number of dimensions, conditions, trials, and covariance smoothness. Error bars represent median $\pm$ SEM across 5 independent runs.
(C1-C4) Same as in B, the operator norm between the true and inferred covariances.
(D) Schematic of QDA classifier.
(E) Using different estimates of covariance, we use QDA to infer the condition label of held-out trials. Error bars represent the standard deviation over 6 independent runs.
(F) Schematic of Fisher Information, FI.
(G) Estimation of FI using grand empirical covariances and Wishart model.
}
\label{fig:synthetic-data}
\end{figure}

% \paragraph{Baselines and Metrics}
We qualitatively and quantitatively compared Wishart processes to common covariance estimation methods used in neural data analysis: Ledoit-Wolf~\cite{Ledoit2004}, empirical covariance (eq.~\ref{eq:empirical-covariance}), grand empirical covariance (eq.~\ref{eq:grand-empirical-covariance-factorization}), and graphical LASSO~\cite{Friedman2007}.
Our primary measure of performance is the log-likelihood that the models assign to held-out data.
Because the baseline methods are not Bayesian, we handicap ourselves by using a single sample from the posterior distribution to calculate held-out log-likelihoods.
We also visualize and qualitatively evaluate contours of the predictive probability density (``covariance ellipses'') in the top two principal component subspace (``PC space'').
For synthetic data, we report the difference between inferred and true covariances measured in the operator norm (Fig.~\ref{fig:synthetic-data}C1-C4).
We observed that generally log-likelihood and operator norm results are aligned. However, since they measure the performance in two different ways it is possible for a model to outperform another in log probability but vice versa on the covariance operator norm.

\subsection{Synthetic Data}

We first validate our approach on simulated data over a periodic 1-dimensional space, with conditions sampled at $C$ equispaced angles on the interval $[0, 2\pi)$. 
The mean and covariance of neural responses were sampled according to the Wishart process generative model (eq.~\ref{eq:gaussian-wishart-priors})
and trials for each condition were conditionally sampled according to eq.~(\ref{eq:observation-model}) from a multivariate Gaussian distribution. An example of generated data is shown in Fig.~\ref{fig:synthetic-data}A1. 
Due to our choice for the condition space, we use periodic kernels both for the Gaussian and Wishart process priors.
For simplicity, we set $\mbLambda(\mbx)$ to the identity matrix both in the data generative and inference models.
To match experimental observations~\cite{Rabinowitz2015}, we simulated low-dimensional noise covariance structure by setting scale matrix to $\mbL \mbL^\top = \mbU \texttt{diag}(s_{1:N}) \mbU^T$ where $\mbU$ is a uniformly sampled orthogonal matrix in $N$ dimensions and $s_1>s_2>\dots>s_N>0$ are logarithmically spaced numbers between 1 and $10^{-5}$.
Except for the number of dimensions, conditions, trials, and Wishart smoothness ($\lambda_{\Sigma}$), which we vary over a range to assess the performance of different models, all other hyperparameters are kept constant ($P=2$, $\gamma=0.001$, $\beta=1$, and $\lambda_{\mu}=1$).
Supplement~\ref{supp:simulation_details} provides further details for these simulations, covering all specific hyperparameter settings for each panel of Fig.~\ref{fig:synthetic-data}.

We fit Wishart process models to these simulated datasets using the procedures described in~\cref{subsection:methods-variational-inference}.
These models qualitatively capture the ground truth covariance structure (Fig.~\ref{fig:synthetic-data}A2-A5) and quantitatively outperform alternative methods across a range of synthetic datasets.
Wishart process inference performs particularly well in high-dimensional regimes relevant to neural data analysis with large $N$, small $K$, and intermediate levels of smoothness across conditions (Fig.~\ref{fig:synthetic-data}B-C).
We also investigated the ability of Wishart process models to improve decoding accuracy on this dataset via QDA, as summarized in \cref{subsection:methods-decoding}.
The Wishart process model outperformed the baseline models across a range of data smoothness suggesting that it captures the boundaries between classes more accurately (Fig.~\ref{fig:synthetic-data}E).
Furthermore, in Fig.~\ref{fig:synthetic-data}G we show that on synthetic data the Wishart process model accurately estimates the FI continuously across conditions.
% Note that we can only validate this analysis on synthetic data where the ground truth FI is known.

\begin{figure}
\includegraphics[width=0.9\linewidth]{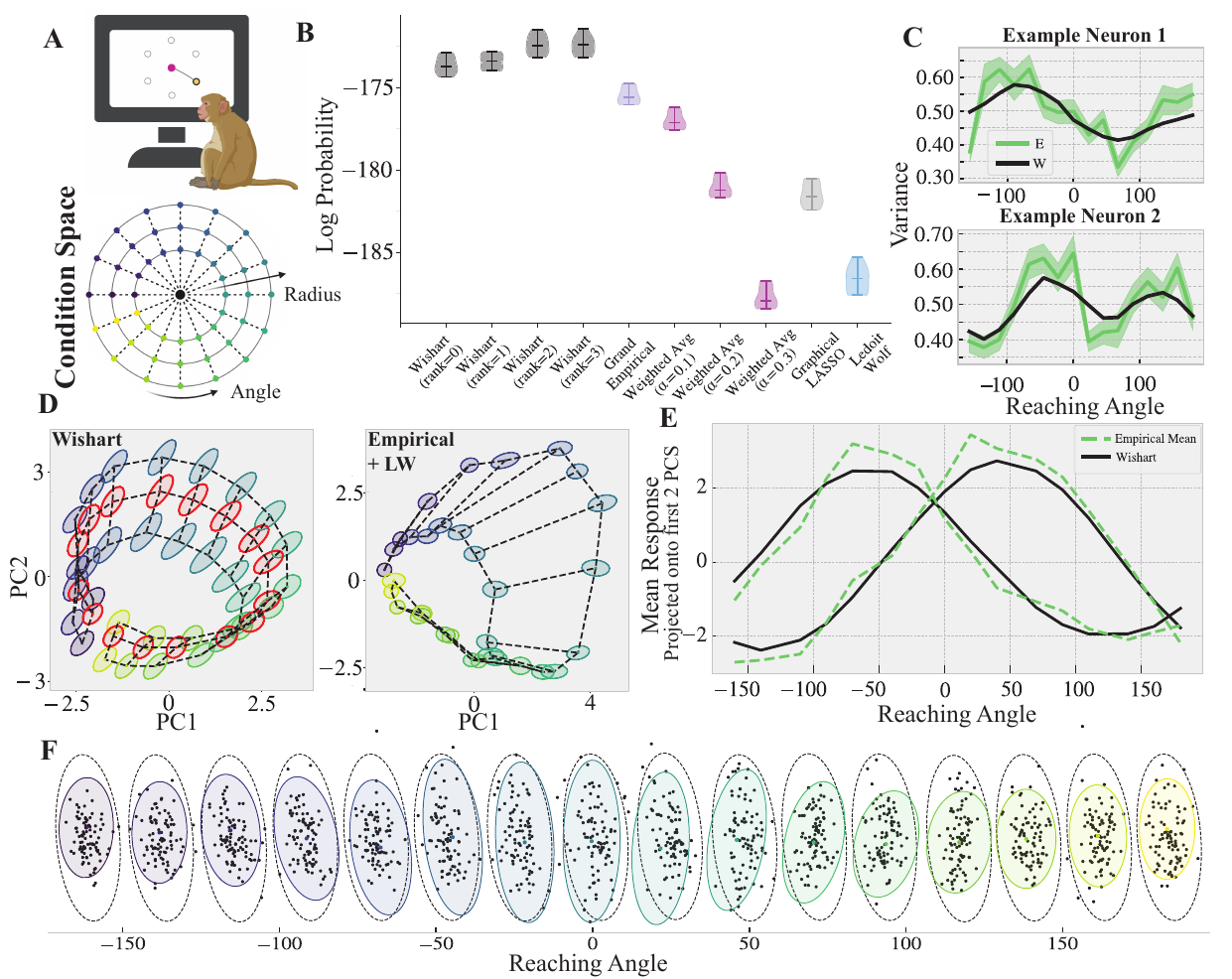} 
\centering
\caption{Results on primate reaching task.
(A) Task schematic (\textit{left}) and set of experimental conditions (\textit{right}).
(B) Log-likelihood of held-out trials across 8 randomized folds. Horizontal lines indicate the maximum, minimum, and mean of the log-likelihood distribution.
(C) Marginal variance of two example neurons across reaching angles, averaged over reach radii (green) and Wishart process model estimate (black).
Shaded green denotes mean $\pm$ SEM.
(D)  Wishart (\textit{left}) and Ledoit-Wolf (\textit{right}) estimates of mean and covariance, projected onto the first two principal components.
To demonstrate the ability of the Wishart Process model to predict covariance in unseen conditions, the middle ring of targets was held out in training (red ellipses). 
(E) Predicted mean and (F) covariance in held-out conditions shown in panel D.
Covariance ellipses, grand-empirical estimates (dashed lines), and samples (black dots) are visualized in the top-2 principal component subspace.
}
\label{fig:monkey-reaching}
\end{figure}

\subsection{Primate Reaching}
\label{subsec:reaching-results}

To study how the brain plans upcoming actions, animals are often trained to prepare voluntary movements to cued locations during a ``delay period'' before executing those movements after a ``go cue'' is delivered~\cite{Rosenbaum1980-it,Wise1985-ao,Galinanes2018}.
During this delay period, neural firing rates in the primary motor cortex (M1) and dorsal premotor cortex (PMd) correlate with features of the upcoming movement (e.g. direction and distance) in nonhuman primates~\cite{Tanji1976-kk,Messier2000-bc}.
Trial-to-trial variability in the delay period has been suggested to correlate with fluctuations in reaction time~\cite{Afshar2011}, but a quantitative account of this variability with dense target configurations is challenging to achieve~\cite{Even-Chen2019-fi}.

We analyzed 200ms of spiking activity prior to the onset of the “go cue” across $N=192$ multiunit channels from both PMd and M1 in a nonhuman primate executing point-to-point hand reaches.\footnote{Data was previously published and made available by \cite{Even-Chen2019-fi}.}
The dataset contains $C=48$ conditions, where targets were organized in three concentric circles at radii of 4cm, 8cm, and 12cm, with 16 equidistant targets around each ring (Fig.~\ref{fig:monkey-reaching}A).
We used a squared exponential kernel (eq.~\ref{eq:kernel-squared-exp}) to model smoothness across reaches of different lengths and a $2\pi$-periodic kernel (eq.~\ref{eq:kernel-periodic}) to model smoothness across reach angles.
Models were fitted on 60\% of the available trials, radial and angular smoothness parameters for both GP and WP kernels were selected on a validation set of 25\% of the data, and log-likelihood scores were generated using the remaining 15\% of trials.
This procedure was repeated over 8 randomized data partitions.
% The performance is quantified over 8 randomized data partitions to ensure that hyperparameters were not overfitted to a particular partition.
%Models were fit to $K=60$ trials per condition and evaluated on a heldout set of $20$ trials per condition.
%We quantified performance over 30 randomized data partitions to ensure that hyperparameters were not overfit to the test set.

The Wishart process model outperforms competing estimators (Fig.~\ref{fig:monkey-reaching}B).
Since $N > K$, the empirical covariance (eq.~\ref{eq:empirical-covariance}) is singular and therefore produces negative infinity log-likelihoods. 
Instead, weighted averages of the grand empirical and empirical covariance were calculated according to $\mbSigma^{\text{WA}}(\mbx_c) = \alpha\mbSigma^{\text{E}}(\mbx_c) + (1-\alpha)\mbSigma^{\text{GE}}$ (WA, E, and GE denote weighted average, empirical, and grand-empirical estimators respectively). 
This baseline is a simplified variant of~\cite{rahim2019population} shrinking the per-condition covariances towards the grand-empirical estimate.
Performance results produced by this baseline are shown in Fig.~\ref{fig:monkey-reaching}B at different values of $\alpha$.

These results suggest that WP is able to capture the smooth change in the covariance across conditions.
Since smooth changes in the covariance imply smooth changes in the marginal variance (diagonal terms in the covariance), we plot the marginal variance of individual neurons as a function of the reaching angle to visualize the smooth change.
% To visualize this, we plot the marginal variance of individual neurons (diagonal terms in the covariance) as a function of the reaching angle.
We see that the variance indeed changes across conditions and that the Wishart process model provides a smooth fit to these noisy variance estimates (Fig.~\ref{fig:monkey-reaching}C).
This smoothness is absent in the predictions of the Ledoit-Wolf and graphical LASSO estimators, while the grand covariance estimate is constant and, in essence, over-smoothed (Supp. Fig. 2).
The out-of-sample log probability results suggest that the smoothness captured by WP is indeed a property that is present in the data and has physiological relevance (Fig.~\ref{fig:monkey-reaching}B).
% To investigate the performance of this estimator in capturing condition specific tuning across reaching angles we compare our method to bootstrapped empirical marginal variance estimated using a leave-one-out strategy (Fig.~\ref{fig:monkey-reaching}C). 
% The Wishart process model smoothly captures this variance tuning.  
% The Wishart process model smoothly captures this variance tuning---for a comparison with the remaining estimation methods see \hl{supplementary materials}.

We also investigated the Wishart model's ability to interpolate covariance estimates for unseen conditions.
Here, the model was retrained on only trials from the inner and outermost rings  (4cm and 12cm).  
The neural response means and covariances around the center ring were then estimated by sampling from the model posterior evaluated at these hold-out condition locations (8cm, all angles). 
% The model continues to outperform the Ledoit-Wolf estimator despite reducing the number of radial conditions to just two (Fig.~\ref{fig:monkey-reaching}D).  
Wishart model estimations perform comparably to empirical mean and grand-empirical covariance estimations calculated on these hold-out conditions (Fig.~\ref{fig:monkey-reaching}E-F).

\subsection{Allen Brain Observatory (Neuropixels)}
\label{subsec:visual-gratings-results}

\begin{figure}
\centering
\includegraphics[width=.95\linewidth]{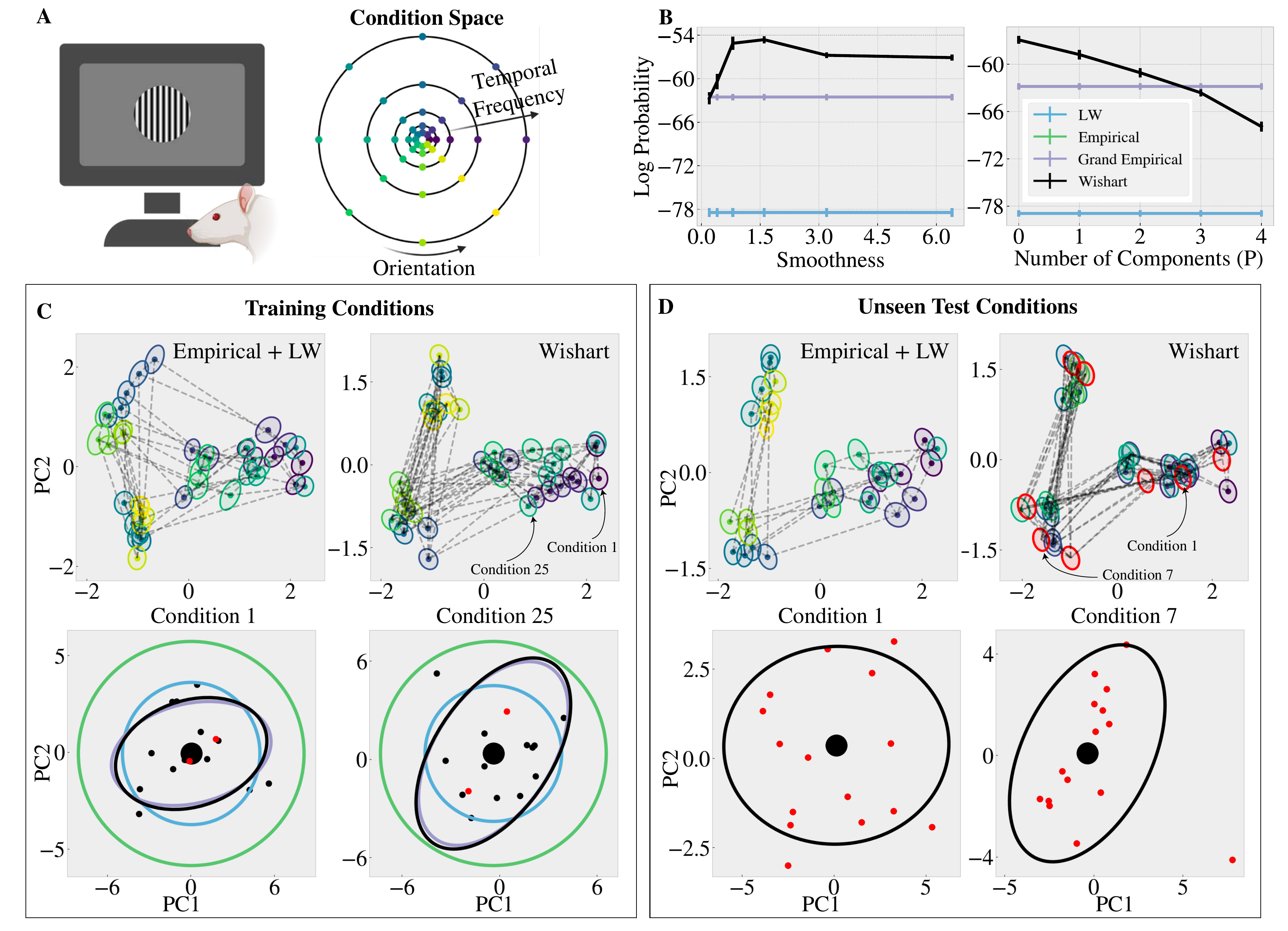} 
\caption{
Results on drifted gratings dataset.
(A) Task schematic (\textit{left}) and set of experimental conditions (\textit{right}). 
% (A) Left: Mouse visual cortex neural spike counts are collected in response to repeated presentation of drifting grating stimuli; Right: Stimuli are presented at different orientations and temporal frequencies parameterized by a 2D condition space. 
% 
(B) Cross-validated log probability as a function of Wishart smoothness  ($\lambda_{\Sigma}$) and the number of components ($P$) achieving their optimal values at $\lambda_{\Sigma} \approx 1.5$ and $P = 0$ and outperforming alternatives.
(C) Top: mean responses and noise covariances of the full set of conditions plotted in the PC space for the Ledoit-Wolf (\textit{left}) vs. Wishart (\textit{right}) estimators (covariances are scaled down by a factor of 100 for visualization
purposes).
%We used empirical mean estimation and LW covariance estimation as the baseline.
Colors are in the increasing order of condition number matching the colors in (A); Bottom: Train and test trials (black and red dots) in two example conditions plotted in the top two PC subspace.
Covariance ellipses are color-coded by the model as in panel B.
(D) Generalizing to unseen conditions: Top: 80\% of the conditions are used for training (shown in colors) and the remaining conditions (shown in red) are withheld. Bottom: Inferred covariances by Wishart Process model for two held-out conditions (as in panel C, bottom). 
Notice that panels C, and D are from two separate analyses. In panel C, we train on all conditions and test on held out trials.
In panel D, we train on a subset of conditions and test on held out conditions (interpolation over stimulus space). Condition 1 was held out only in the second analysis.
}
% \vspace{-1em}
\label{fig:allen-brain-data}
\end{figure}

Perceptual discrimination is a longstanding area of research in psychology and neuroscience, dating back to work in the early 1800s by Weber and Fechner~\cite{fechner1860elemente}.
A classic paradigm requires subjects to discriminate between pairs of oriented visual gratings, which humans and nonhuman primates accomplish with high precision~\cite{Webster1990,Burr1991-wv,Vogels1990-nn}.
Discrimination thresholds on this task are thought to be related to the scale and orientation of noise in neural representations~\cite{Moreno2014} (Fig.~\ref{fig:schematics}A), and recent works have examined this hypothesis by simultaneously recording large populations of neurons in rodent models~\cite{Rumyantsev2020-ar,Kafashan2021,Stringer2021}.
The resulting analyses are delicate and require a large number of trials.

To investigate, we analyzed simultaneously recorded responses in mouse primary visual cortex to drifting visual gratings in the Allen Brain Observatory (Visual Coding: Neuropixels dataset).\footnote{\url{https://portal.brain-map.org/explore/circuits/visual-coding-neuropixels}}
We analyzed a session (session-id = 756029989) with 81 neurons.
Spike count responses in a 200 ms window from the stimulus onset were recorded across 4 grating orientations, 2 directions, and 5 temporal frequencies as illustrated in Fig.~\ref{fig:allen-brain-data}A.
%
% We respectively used periodic and squared exponential kernels (eq.~\ref{eq:kernel-periodic} and eq.~\ref{eq:kernel-squared-exp}) for the orientation, direction, and temporal frequency axes of the condition space.
We used periodic kernels for the orientation and direction axes (eq.~\ref{eq:kernel-periodic}), and a squared exponential kernel (eq.~\ref{eq:kernel-squared-exp}) for the temporal frequency axis of the condition space.
For each of the resulting 40 conditions, 15 trials are collected, which we split into 13 train vs. 2 test trials.
For choosing the number of components ($P$) and the smoothness of the Wishart kernel ($\lambda_{\Sigma}$), we quantified the cross-validated log probability of test trials over 10 randomized folds.
As depicted in Fig.~\ref{fig:allen-brain-data}B the optimal performance of the Wishart model is achieved for $\lambda_{\Sigma} \approx 1.5$ and $P=0$. 

This selection of the model parameters allows us to capture the structure that is shared among all conditions (note the similarity between Wishart and grand empirical covariance in Fig.~\ref{fig:allen-brain-data}C), while representing the condition-specific changes in the covariance structure (note the smooth change of matrix orientation across conditions in the top-right plot of Fig.~\ref{fig:allen-brain-data}C).
Finally, we present covariance interpolation results across unseen conditions where we fit the model according to 32 randomly selected train conditions (colored ellipses in Fig~\ref{fig:allen-brain-data}D top-left) and leave out the complete set of trials for the remaining 8 conditions (red ellipses in Fig~\ref{fig:allen-brain-data}D top-right). The interpolated means and covariances for two example test conditions present a strong agreement with their corresponding held-out trials (Fig~\ref{fig:allen-brain-data}D bottom row).

\section{Conclusion}
We proposed a method to estimate covariance across large populations of co-recorded neurons, which is widely regarded as a challenging problem~\cite{Williams2021-single-trial-limit,Yatsenko2015-vy} with important implications for theories of neural coding~\cite{Moreno2014,Panzeri2022,averbeck2006neural}.
Specifically, we formulated probabilistic Gaussian and Wishart process models~\cite{Wilson2011} to model smoothly changing covariance structure across continuously parameterized experimental conditions.
We found this approach outperforms off-the-shelf covariance estimators that do not exploit this smoothness (e.g. Ledoit-Wolf) as well as the grand covariance estimator (eq.~\ref{eq:grand-empirical-covariance-factorization}) which is infinitely smooth (i.e. does not allow for condition-dependent covariance).
We investigated two task modalities (vision and motor) and model organisms (mouse and nonhuman primate).
However, we expect these modeling principles will be even more broadly applicable, including to other sensory modalities with smoothy parameterized stimuli (e.g. acoustic tones or odorant mixtures) and in neural circuits supporting navigation~\cite{Peyrache2015-ov,Hazon2022-ff}.

We believe our work also provides an important conceptual advance.
Given the limited duration of experiments, it may seem that there is a strict tradeoff between accurately estimating neural responses across many experimental conditions and estimating trial-by-trial noise statistics in any fixed condition.
Our results show that this tradeoff is not so simple.
Indeed, if neural responses are sufficiently smooth across conditions, the Wishart process model could predict neural covariance accurately by densely sampling conditions with only a single trial per condition (e.g. visual gratings at random orientations, as done in \cite{Stringer2021}).
Intuitively, the model works by pooling information across nearby conditions, and so can even predict covariance in entirely unseen conditions (as demonstrated in Figs.~\ref{fig:monkey-reaching}D and \ref{fig:allen-brain-data}D).
Thus, the limiting factor in estimating noise covariance is not the number of trials per condition, but the smoothness of the neural responses over the space of experimental conditions.

We conclude by discussing the current limitations of our model and opportunities for future work.
First, the model requires us to explicitly specify kernel functions ($k_\mu$, $k_\Sigma$) over the space of experimental conditions.
We studied simple paradigms where a reasonable parametric form of this kernel could be specified and the hyperparameters could be fine-tuned through cross-validation.
Extending the model to more complex stimulus sets (e.g. to naturalistic image stimuli) would be of great interest, and likely require more sophisticated approaches for learning kernels and feature representations of these stimulus sets.
Second, we assumed that trial-to-trial noise followed a Gaussian distribution.
A deeper investigation of noise models, including multivariate extensions of Poisson distributions~\cite{Inouye2017,Rabinowitz2015,Henaff2021-xa}, is warranted.
Finally, we utilized a basic variational inference procedure to approximate the posterior distribution over neural covariance.
While fully representing the posterior is intractable (as is typically the case), more sophisticated optimization techniques may further benefit performance~\cite{burda2015importance}.
Nonetheless, our results show that Wishart process models are performant even in the absence of the aforementioned extensions.
These models outperform common practices in neuroscience and open the door to new analyses, such as the prediction of noise covariance in held-out conditions.

\section*{Acknowledgements}

We thank Saurabh Vyas (Columbia) and Eric Trautmann (Columbia) for helpful discussions and for providing data from the nonhuman primate motor cortex. We also thank Sam Zheng (NYU) and Sarah Harvey (NYU) for their feedback on the manuscript.

\printbibliography

@ARTICLE{Johnstone2018,
  author={Johnstone, Iain M. and Paul, Debashis},
  journal={Proceedings of the IEEE}, 
  title={PCA in High Dimensions: An Orientation}, 
  year={2018},
  volume={106},
  number={8},
  pages={1277-1292},
  doi={10.1109/JPROC.2018.2846730}}

@ARTICLE{Galinanes2018,
  title    = "Directional Reaching for Water as a {Cortex-Dependent} Behavioral
              Framework for Mice",
  author   = "Gali{\~n}anes, Gregorio Luis and Bonardi, Claudia and Huber,
              Daniel",
  journal  = "Cell Rep.",
  volume   =  22,
  number   =  10,
  pages    = "2767--2783",
  month    =  mar,
  year     =  2018,
  language = "en"
}

@ARTICLE{Tanji1976-kk,
  title    = "Anticipatory activity of motor cortex neurons in relation to
              direction of an intended movement",
  author   = "Tanji, J and Evarts, E V",
  journal  = "J. Neurophysiol.",
  volume   =  39,
  number   =  5,
  pages    = "1062--1068",
  month    =  sep,
  year     =  1976,
  language = "en"
}

@ARTICLE{Messier2000-bc,
  title    = "Covariation of primate dorsal premotor cell activity with
              direction and amplitude during a memorized-delay reaching task",
  author   = "Messier, J and Kalaska, J F",
  journal  = "J. Neurophysiol.",
  volume   =  84,
  number   =  1,
  pages    = "152--165",
  month    =  jul,
  year     =  2000,
  language = "en"
}

@article{Afshar2011,
title = {Single-Trial Neural Correlates of Arm Movement Preparation},
journal = {Neuron},
volume = {71},
number = {3},
pages = {555-564},
year = {2011},
issn = {0896-6273},
doi = {https://doi.org/10.1016/j.neuron.2011.05.047},
author = {Afsheen Afshar and Gopal Santhanam and Byron M. Yu and Stephen I. Ryu and Maneesh Sahani and Krishna V. Shenoy}
}

@ARTICLE{Moreno2014,
  title    = "Information-limiting correlations",
  author   = "Moreno-Bote, Rub{\'e}n and Beck, Jeffrey and Kanitscheider,
              Ingmar and Pitkow, Xaq and Latham, Peter and Pouget, Alexandre",
  journal  = "Nat. Neurosci.",
  volume   =  17,
  number   =  10,
  pages    = "1410--1417",
  month    =  oct,
  year     =  2014,
  language = "en"
}

@article{averbeck2006neural,
  title={Neural correlations, population coding and computation},
  author={Averbeck, Bruno B and Latham, Peter E and Pouget, Alexandre},
  journal={Nature reviews neuroscience},
  volume={7},
  number={5},
  pages={358--366},
  year={2006},
  publisher={Nature Publishing Group UK London}
}

@book{murphy2012machine,
  title={Machine learning: a probabilistic perspective},
  author={Murphy, Kevin P},
  year={2012},
  publisher={MIT press}
}

@ARTICLE{Abbott1999-tb,
  title    = "The effect of correlated variability on the accuracy of a
              population code",
  author   = "Abbott, L F and Dayan, P",
  journal  = "Neural Comput.",
  volume   =  11,
  number   =  1,
  pages    = "91--101",
  month    =  jan,
  year     =  1999,
  language = "en"
}

@ARTICLE{Yatsenko2015-vy,
  title     = "Improved estimation and interpretation of correlations in neural
               circuits",
  author    = "Yatsenko, Dimitri and Josi{\'c}, Kre{\v s}imir and Ecker,
               Alexander S and Froudarakis, Emmanouil and Cotton, R James and
               Tolias, Andreas S",
  journal   = "PLoS Comput. Biol.",
  publisher = "journals.plos.org",
  volume    =  11,
  number    =  3,
  pages     = "e1004083",
  month     =  mar,
  year      =  2015,
  language  = "en"
}

@ARTICLE{Gourieroux2009-gp,
  title    = "The Wishart Autoregressive process of multivariate stochastic
              volatility",
  author   = "Gourieroux, C and Jasiak, J and Sufana, R",
  journal  = "J. Econom.",
  volume   =  150,
  number   =  2,
  pages    = "167--181",
  month    =  jun,
  year     =  2009
}

@ARTICLE{Shadlen1996-dr,
  title    = "A computational analysis of the relationship between neuronal and
              behavioral responses to visual motion",
  author   = "Shadlen, M N and Britten, K H and Newsome, W T and Movshon, J A",
  journal  = "J. Neurosci.",
  volume   =  16,
  number   =  4,
  pages    = "1486--1510",
  month    =  feb,
  year     =  1996,
  language = "en"
}

@ARTICLE{Zohary1994-ns,
  title    = "Correlated neuronal discharge rate and its implications for
              psychophysical performance",
  author   = "Zohary, E and Shadlen, M N and Newsome, W T",
  journal  = "Nature",
  volume   =  370,
  number   =  6485,
  pages    = "140--143",
  month    =  jul,
  year     =  1994,
  language = "en"
}

@inproceedings{keeley2020efficient,
  title={Efficient non-conjugate Gaussian process factor models for spike count data using polynomial approximations},
  author={Keeley, Stephen and Zoltowski, David and Yu, Yiyi and Smith, Spencer and Pillow, Jonathan},
  booktitle={International Conference on Machine Learning},
  pages={5177--5186},
  year={2020},
  organization={PMLR}
}

@inproceedings{burda2015importance,
  title={Importance weighted autoencoders},
  author={Burda, Yuri and Grosse, Roger and Salakhutdinov, Ruslan},
  journal={International Conference on Learning Representations},
  year={2015}
}

@article{goris2014partitioning,
  title={Partitioning neuronal variability},
  author={Goris, Robbe LT and Movshon, J Anthony and Simoncelli, Eero P},
  journal={Nature neuroscience},
  volume={17},
  number={6},
  pages={858--865},
  year={2014},
  publisher={Nature Publishing Group US New York}
}

@ARTICLE{Buzsaki2014-ea,
  title    = "The log-dynamic brain: how skewed distributions affect network
              operations",
  author   = "Buzs{\'a}ki, Gy{\"o}rgy and Mizuseki, Kenji",
  journal  = "Nat. Rev. Neurosci.",
  volume   =  15,
  number   =  4,
  pages    = "264--278",
  month    =  apr,
  year     =  2014,
  language = "en"
}

@ARTICLE{Cohen2011-ke,
  title    = "Measuring and interpreting neuronal correlations",
  author   = "Cohen, Marlene R and Kohn, Adam",
  journal  = "Nat. Neurosci.",
  volume   =  14,
  number   =  7,
  pages    = "811--819",
  month    =  jun,
  year     =  2011,
  language = "en"
}

@ARTICLE{Panzeri2022,
  title    = "The structures and functions of correlations in neural population
              codes",
  author   = "Panzeri, Stefano and Moroni, Monica and Safaai, Houman and
              Harvey, Christopher D",
  journal  = "Nat. Rev. Neurosci.",
  volume   =  23,
  number   =  9,
  pages    = "551--567",
  month    =  sep,
  year     =  2022,
  language = "en"
}

@ARTICLE{Zhu2022-ey,
  title    = "A deep learning framework for inference of single-trial neural
              population dynamics from calcium imaging with subframe temporal
              resolution",
  author   = "Zhu, Feng and Grier, Harrison A and Tandon, Raghav and Cai,
              Changjia and Agarwal, Anjali and Giovannucci, Andrea and Kaufman,
              Matthew T and Pandarinath, Chethan",
  journal  = "Nat. Neurosci.",
  volume   =  25,
  number   =  12,
  pages    = "1724--1734",
  month    =  dec,
  year     =  2022,
  language = "en"
}

@ARTICLE{Inouye2017,
  title    = "A Review of Multivariate Distributions for Count Data Derived
              from the Poisson Distribution",
  author   = "Inouye, David and Yang, Eunho and Allen, Genevera and Ravikumar,
              Pradeep",
  journal  = "Wiley Interdiscip. Rev. Comput. Stat.",
  volume   =  9,
  number   =  3,
  month    =  mar,
  year     =  2017,
  language = "en"
}

@inproceedings{Keeley2020identifying,
 author = {Keeley, Stephen and Aoi, Mikio and Yu, Yiyi and Smith, Spencer and Pillow, Jonathan W},
 booktitle = {Advances in Neural Information Processing Systems},
 editor = {H. Larochelle and M. Ranzato and R. Hadsell and M.F. Balcan and H. Lin},
 pages = {13795--13805},
 publisher = {Curran Associates, Inc.},
 title = {Identifying signal and noise structure in neural population activity with Gaussian process factor models},
 url = {https://proceedings.neurips.cc/paper_files/paper/2020/file/9eed867b73ab1eab60583c9d4a789b1b-Paper.pdf},
 volume = {33},
 year = {2020}
}

@inproceedings{Rutten2020,
 author = {Rutten, Virginia and Bernacchia, Alberto and Sahani, Maneesh and Hennequin, Guillaume},
 booktitle = {Advances in Neural Information Processing Systems},
 editor = {H. Larochelle and M. Ranzato and R. Hadsell and M.F. Balcan and H. Lin},
 pages = {9622--9632},
 publisher = {Curran Associates, Inc.},
 title = {Non-reversible Gaussian processes for identifying latent dynamical structure in neural data},
 url = {https://proceedings.neurips.cc/paper_files/paper/2020/file/6d79e030371e47e6231337805a7a2685-Paper.pdf},
 volume = {33},
 year = {2020}
}

@article{ober2021variational,
  title={A variational approximate posterior for the deep Wishart process},
  author={Ober, Sebastian and Aitchison, Laurence},
  journal={Advances in Neural Information Processing Systems},
  volume={34},
  pages={6567--6579},
  year={2021}
}

@article{Stringer2021,
title = {High-precision coding in visual cortex},
journal = {Cell},
volume = {184},
number = {10},
pages = {2767-2778.e15},
year = {2021},
issn = {0092-8674},
doi = {https://doi.org/10.1016/j.cell.2021.03.042},
url = {https://www.sciencedirect.com/science/article/pii/S0092867421003731},
author = {Carsen Stringer and Michalis Michaelos and Dmitri Tsyboulski and Sarah E. Lindo and Marius Pachitariu},
}

@ARTICLE{Walther2016-se,
  title    = "Reliability of dissimilarity measures for multi-voxel pattern
              analysis",
  author   = "Walther, Alexander and Nili, Hamed and Ejaz, Naveed and Alink,
              Arjen and Kriegeskorte, Nikolaus and Diedrichsen, J{\"o}rn",
  journal  = "Neuroimage",
  volume   =  137,
  pages    = "188--200",
  month    =  aug,
  year     =  2016,
  language = "en"
}

@ARTICLE{Henaff2021-xa,
  title    = "Primary visual cortex straightens natural video trajectories",
  author   = "H{\'e}naff, Olivier J and Bai, Yoon and Charlton, Julie A and
              Nauhaus, Ian and Simoncelli, Eero P and Goris, Robbe L T",
  journal  = "Nat. Commun.",
  volume   =  12,
  number   =  1,
  pages    = "5982",
  month    =  oct,
  year     =  2021,
  language = "en"
}

@article{Li2021,
    title = {Joint representation of working memory and uncertainty in human cortex},
    journal = {Neuron},
    volume = {109},
    number = {22},
    pages = {3699-3712.e6},
    year = {2021},
    issn = {0896-6273},
    doi = {https://doi.org/10.1016/j.neuron.2021.08.022},
    url = {https://www.sciencedirect.com/science/article/pii/S089662732100619X},
    author = {Hsin-Hung Li and Thomas C. Sprague and Aspen H. Yoo and Wei Ji Ma and Clayton E. Curtis},
}

@ARTICLE{Cai2019-pi,
  title    = "Representational structure or task structure? Bias in neural
              representational similarity analysis and a Bayesian method for
              reducing bias",
  author   = "Cai, Ming Bo and Schuck, Nicolas W and Pillow, Jonathan W and
              Niv, Yael",
  journal  = "PLoS Comput. Biol.",
  volume   =  15,
  number   =  5,
  pages    = "e1006299",
  month    =  may,
  year     =  2019,
  language = "en"
}

@Article{Kafashan2021,
author={Kafashan, MohammadMehdi
and Jaffe, Anna W.
and Chettih, Selmaan N.
and Nogueira, Ramon
and Arandia-Romero, I{\~{n}}igo
and Harvey, Christopher D.
and Moreno-Bote, Rub{\'e}n
and Drugowitsch, Jan},
title={Scaling of sensory information in large neural populations shows signatures of information-limiting correlations},
journal={Nature Communications},
year={2021},
month={Jan},
day={20},
volume={12},
number={1},
pages={473},
issn={2041-1723},
doi={10.1038/s41467-020-20722-y},
url={https://doi.org/10.1038/s41467-020-20722-y}
}

@ARTICLE{Vogels1990-nn,
  title    = "How well do response changes of striate neurons signal
              differences in orientation: a study in the discriminating monkey",
  author   = "Vogels, R and Orban, G A",
  journal  = "J. Neurosci.",
  volume   =  10,
  number   =  11,
  pages    = "3543--3558",
  month    =  nov,
  year     =  1990,
  language = "en"
}

@ARTICLE{Burr1991-wv,
  title    = "Orientation discrimination depends on spatial frequency",
  author   = "Burr, D C and Wijesundra, S A",
  journal  = "Vision Res.",
  volume   =  31,
  number   = "7-8",
  pages    = "1449--1452",
  year     =  1991,
  language = "en"
}

@article{Webster1990,
    author = {Michael A. Webster and Karen K. De Valois and Eugene Switkes},
    journal = {J. Opt. Soc. Am. A},
    number = {6},
    pages = {1034--1049},
    publisher = {Optica Publishing Group},
    title = {Orientation and spatial-frequency discrimination for luminance and chromatic gratings},
    volume = {7},
    month = {Jun},
    year = {1990},
    url = {https://opg.optica.org/josaa/abstract.cfm?URI=josaa-7-6-1034},
    doi = {10.1364/JOSAA.7.001034},
}

@book{fechner1860elemente,
  title={Elemente der psychophysik},
  author={Fechner, Gustav Theodor},
  volume={2},
  year={1860},
  publisher={Breitkopf u. H{\"a}rtel}
}

@article{numpyro,
  title={Composable Effects for Flexible and Accelerated Probabilistic Programming in NumPyro},
  author={Phan, Du and Pradhan, Neeraj and Jankowiak, Martin},
  journal={arXiv preprint arXiv:1912.11554},
  year={2019}
}

@book{gelman2013bayesian,
  title={Bayesian data analysis (Third Edition)},
  author={Gelman, Andrew and Carlin, John B and Stern, Hal S and Dunson, David B and Vehtari, Aki and Rubin, Donald B},
  year={2013},
  publisher={CRC press}
}

@inproceedings{Heaukulani2019,
    author = {Heaukulani, Creighton and van der Wilk, Mark},
    booktitle = {Advances in Neural Information Processing Systems},
    editor = {H. Wallach and H. Larochelle and A. Beygelzimer and F. d\textquotesingle Alch\'{e}-Buc and E. Fox and R. Garnett},
    pages = {},
    publisher = {Curran Associates, Inc.},
    title = {Scalable Bayesian dynamic covariance modeling with variational Wishart and inverse Wishart processes},
    url = {https://proceedings.neurips.cc/paper_files/paper/2019/file/5b168fdba5ee5ea262cc2d4c0b457697-Paper.pdf},
    volume = {32},
    year = {2019}
}

@inproceedings{Wilson2011,
  title={Generalised Wishart processes},
  author={Wilson, Andrew Gordon and Ghahramani, Zoubin},
  booktitle={Proceedings of the 27th Conference on Uncertainty in Artificial Intelligence, UAI 2011},
  pages={736--744},
  year={2011},
  organization={AUAI Press}
}

@book{Pourahmadi2013,
  title={High-dimensional covariance estimation: with high-dimensional data},
  author={Pourahmadi, Mohsen},
  volume={882},
  year={2013},
  publisher={John Wiley \& Sons}
}

@inproceedings{Duong2023,
  author = {Lyndon R. Duong and Jingyang Zhou and Josue Nassar and Jules Berman and Jeroen Olieslagers and Alex H. Williams},
  title = {Representational dissimilarity metric spaces for stochastic neural networks},
  year = {2023},
  booktitle = {International Conference on Learning Representations},
}

@ARTICLE{Wise1985-ao,
  title    = "The primate premotor cortex: past, present, and preparatory",
  author   = "Wise, S P",
  journal  = "Annu. Rev. Neurosci.",
  volume   =  8,
  pages    = "1--19",
  year     =  1985,
  language = "en"
}

@ARTICLE{Rosenbaum1980-it,
  title    = "Human movement initiation: Specification of arm, direction, and
              extent",
  author   = "Rosenbaum, David A",
  journal  = "J. Exp. Psychol. Gen.",
  volume   =  109,
  number   =  4,
  pages    = "444--474",
  month    =  dec,
  year     =  1980
}

@inproceedings{Dapello2021,
 author = {Dapello, Joel and Feather, Jenelle and Le, Hang and Marques, Tiago and Cox, David and McDermott, Josh and DiCarlo, James J and Chung, Sueyeon},
 booktitle = {Advances in Neural Information Processing Systems},
 editor = {M. Ranzato and A. Beygelzimer and Y. Dauphin and P.S. Liang and J. Wortman Vaughan},
 pages = {15595--15607},
 publisher = {Curran Associates, Inc.},
 title = {Neural Population Geometry Reveals the Role of Stochasticity in Robust Perception},
 url = {https://proceedings.neurips.cc/paper_files/paper/2021/file/8383f931b0cefcc631f070480ef340e1-Paper.pdf},
 volume = {34},
 year = {2021}
}

@article{Friedman2007,
    author = {Friedman, Jerome and Hastie, Trevor and Tibshirani, Robert},
    title = "{Sparse inverse covariance estimation with the graphical lasso}",
    journal = {Biostatistics},
    volume = {9},
    number = {3},
    pages = {432-441},
    year = {2007},
    month = {12},
    issn = {1465-4644},
    doi = {10.1093/biostatistics/kxm045},
    url = {https://doi.org/10.1093/biostatistics/kxm045},
    eprint = {https://academic.oup.com/biostatistics/article-pdf/9/3/432/17742149/kxm045.pdf},
}

@article{Cohen2008,
	author = {Marlene R. Cohen and William T. Newsome},
	doi = {https://doi.org/10.1016/j.neuron.2008.08.007},
	issn = {0896-6273},
	journal = {Neuron},
	number = {1},
	pages = {162-173},
	title = {Context-Dependent Changes in Functional Circuitry in Visual Area MT},
	volume = {60},
	year = {2008},
}

@ARTICLE{Blei2017-mm,
  title     = "Variational Inference: A Review for Statisticians",
  author    = "Blei, David M and Kucukelbir, Alp and McAuliffe, Jon D",
  journal   = "J. Am. Stat. Assoc.",
  publisher = "Taylor \& Francis",
  volume    =  112,
  number    =  518,
  pages     = "859--877",
  month     =  apr,
  year      =  2017
}

@ARTICLE{Kohn2005,
  title    = "Stimulus dependence of neuronal correlation in primary visual
              cortex of the macaque",
  author   = "Kohn, Adam and Smith, Matthew A",
  journal  = "J. Neurosci.",
  volume   =  25,
  number   =  14,
  pages    = "3661--3673",
  month    =  apr,
  year     =  2005,
  language = "en"
}

@article{Alvarez2013,
    author = {Adrian Ponce-Alvarez  and Alexander Thiele  and Thomas D. Albright  and Gene R. Stoner  and Gustavo Deco },
    title = {Stimulus-dependent variability and noise correlations in cortical MT neurons},
    journal = {Proceedings of the National Academy of Sciences},
    volume = {110},
    number = {32},
    pages = {13162-13167},
    year = {2013},
    doi = {10.1073/pnas.1300098110},
}

@article{Rabinowitz2015,
	article_type = {journal},
	author = {Rabinowitz, Neil C and Goris, Robbe L and Cohen, Marlene and Simoncelli, Eero P},
	citation = {eLife 2015;4:e08998},
	doi = {10.7554/eLife.08998},
	editor = {Carandini, Matteo},
	issn = {2050-084X},
	journal = {eLife},
	month = {nov},
	pages = {e08998},
	pub_date = {2015-11-02},
	publisher = {eLife Sciences Publications, Ltd},
	title = {Attention stabilizes the shared gain of V4 populations},
	url = {https://doi.org/10.7554/eLife.08998},
	volume = 4,
	year = 2015
 }

@article{Ledoit2004,
title = {A well-conditioned estimator for large-dimensional covariance matrices},
journal = {Journal of Multivariate Analysis},
volume = {88},
number = {2},
pages = {365-411},
year = {2004},
issn = {0047-259X},
doi = {https://doi.org/10.1016/S0047-259X(03)00096-4},
url = {https://www.sciencedirect.com/science/article/pii/S0047259X03000964},
author = {Olivier Ledoit and Michael Wolf},
}

@inproceedings{Wu2017,
	author = {Wu, Anqi and Roy, Nicholas A. and Keeley, Stephen and Pillow, Jonathan W},
	booktitle = {Advances in Neural Information Processing Systems},
	editor = {I. Guyon and U. Von Luxburg and S. Bengio and H. Wallach and R. Fergus and S. Vishwanathan and R. Garnett},
	publisher = {Curran Associates, Inc.},
	title = {Gaussian process based nonlinear latent structure discovery in multivariate spike train data},
	volume = {30},
	year = {2017},
}

@article{Mackay1998,
  title={Introduction to Gaussian processes},
  author={MacKay, David JC and others},
  journal={NATO ASI series F computer and systems sciences},
  volume={168},
  pages={133--166},
  year={1998},
  publisher={Citeseer}
}

@inproceedings{Yu2008,
	author = {Yu, Byron M and Cunningham, John P and Santhanam, Gopal and Ryu, Stephen and Shenoy, Krishna V and Sahani, Maneesh},
	booktitle = {Advances in Neural Information Processing Systems},
	editor = {D. Koller and D. Schuurmans and Y. Bengio and L. Bottou},
	publisher = {Curran Associates, Inc.},
	title = {Gaussian-process factor analysis for low-dimensional single-trial analysis of neural population activity},
	volume = {21},
	year = {2008},
}

@INPROCEEDINGS{Duncker2019-zm,
  title     = "Learning interpretable continuous-time models of latent
               stochastic dynamical systems",
  booktitle = "Proceedings of the 36th International Conference on Machine
               Learning",
  author    = "Duncker, Lea and Bohner, Gergo and Boussard, Julien and Sahani,
               Maneesh",
  editor    = "Chaudhuri, Kamalika and Salakhutdinov, Ruslan",
  publisher = "PMLR",
  volume    =  97,
  pages     = "1726--1734",
  series    = "Proceedings of Machine Learning Research",
  year      =  2019,
  address   = "Long Beach, California, USA"
}

@book{Williams2006-gps,
  title={Gaussian processes for machine learning},
  author={Williams, Christopher KI and Rasmussen, Carl Edward},
  volume={2},
  number={3},
  year={2006},
  publisher={MIT press Cambridge, MA}
}

@BOOK{Scholkopf2001,
  title     = "Learning with kernels",
  author    = "Scholkopf, Bernhard and Smola, Alexander J",
  publisher = "MIT Press",
  series    = "Adaptive Computation and Machine Learning series",
  month     =  dec,
  year      =  2001,
  address   = "London, England",
  language  = "en"
}

@article{Hofmann2008,
	author = {Thomas Hofmann and Bernhard Sch{\"o}lkopf and Alexander J. Smola},
	doi = {10.1214/009053607000000677},
	journal = {The Annals of Statistics},
	number = {3},
	pages = {1171 -- 1220},
	publisher = {Institute of Mathematical Statistics},
	title = {{Kernel methods in machine learning}},
	volume = {36},
	year = {2008},
 }

@ARTICLE{Even-Chen2019-fi,
  title     = "Structure and variability of delay activity in premotor cortex",
  author    = "Even-Chen, Nir and Sheffer, Blue and Vyas, Saurabh and Ryu,
               Stephen I and Shenoy, Krishna V",
  journal   = "PLoS Comput. Biol.",
  publisher = "journals.plos.org",
  volume    =  15,
  number    =  2,
  pages     = "e1006808",
  month     =  feb,
  year      =  2019,
  language  = "en"
}

@ARTICLE{Peyrache2015-ov,
  title    = "Internally organized mechanisms of the head direction sense",
  author   = "Peyrache, Adrien and Lacroix, Marie M and Petersen, Peter C and
              Buzs{\'a}ki, Gy{\"o}rgy",
  journal  = "Nat. Neurosci.",
  volume   =  18,
  number   =  4,
  pages    = "569--575",
  month    =  apr,
  year     =  2015,
  language = "en"
}

@ARTICLE{Hazon2022-ff,
  title    = "Noise correlations in neural ensemble activity limit the accuracy of hippocampal spatial representations",
  author   = "Hazon, Omer and Minces, Victor H and Tom{\`a}s, David P and
              Ganguli, Surya and Schnitzer, Mark J and Jercog, Pablo E",
  journal  = "Nat. Commun.",
  volume   =  13,
  number   =  1,
  pages    = "4276",
  month    =  jul,
  year     =  2022,
  language = "en"
}

@ARTICLE{Merzenich1975-ko,
  title    = "Representation of cochlea within primary auditory cortex in the
              cat",
  author   = "Merzenich, M M and Knight, P L and Roth, G L",
  journal  = "J. Neurophysiol.",
  volume   =  38,
  number   =  2,
  pages    = "231--249",
  month    =  mar,
  year     =  1975,
  language = "en"
}

@ARTICLE{Kudryavitskaya2021-ky,
  title     = "Flexible categorization in the mouse olfactory bulb",
  author    = "Kudryavitskaya, Elena and Marom, Eran and Shani-Narkiss, Haran
               and Pash, David and Mizrahi, Adi",
  journal   = "Curr. Biol.",
  publisher = "Elsevier",
  volume    =  31,
  number    =  8,
  pages     = "1616--1631.e4",
  month     =  apr,
  year      =  2021,
  language  = "en"
}

@ARTICLE{Rumyantsev2020-ar,
  title    = "Fundamental bounds on the fidelity of sensory cortical coding",
  author   = "Rumyantsev, Oleg I and Lecoq, J{\'e}r{\^o}me A and Hernandez,
              Oscar and Zhang, Yanping and Savall, Joan and Chrapkiewicz,
              Rados{\l}aw and Li, Jane and Zeng, Hongkui and Ganguli, Surya and
              Schnitzer, Mark J",
  journal  = "Nature",
  volume   =  580,
  number   =  7801,
  pages    = "100--105",
  month    =  apr,
  year     =  2020,
  language = "en"
}

@article{Williams2021-single-trial-limit,
	author = {Alex H. Williams and Scott W. Linderman},
	doi = {https://doi.org/10.1016/j.conb.2021.10.008},
	issn = {0959-4388},
	journal = {Current Opinion in Neurobiology},
	note = {Computational Neuroscience},
	pages = {193-205},
	title = {Statistical neuroscience in the single trial limit},
	url = {https://www.sciencedirect.com/science/article/pii/S0959438821001203},
	volume = {70},
	year = {2021}
}

@article{rahim2019population,
  title={Population shrinkage of covariance (PoSCE) for better individual brain functional-connectivity estimation},
  author={Rahim, Mehdi and Thirion, Bertrand and Varoquaux, Ga{\"e}l},
  journal={Medical image analysis},
  volume={54},
  pages={138--148},
  year={2019},
  publisher={Elsevier}
}

@article{pagan2016neural,
  title={Neural quadratic discriminant analysis: Nonlinear decoding with V1-like computation},
  author={Pagan, Marino and Simoncelli, Eero P and Rust, Nicole C},
  journal={Neural computation},
  volume={28},
  number={11},
  pages={2291--2319},
  year={2016},
  publisher={MIT Press One Rogers Street, Cambridge, MA 02142-1209, USA journals-info~…}
}

\clearpage
\setcounter{figure}{4}

\begin{appendices}

\section{Supplemental Figures}

\begin{figure}[h]
\includegraphics[width=\linewidth]{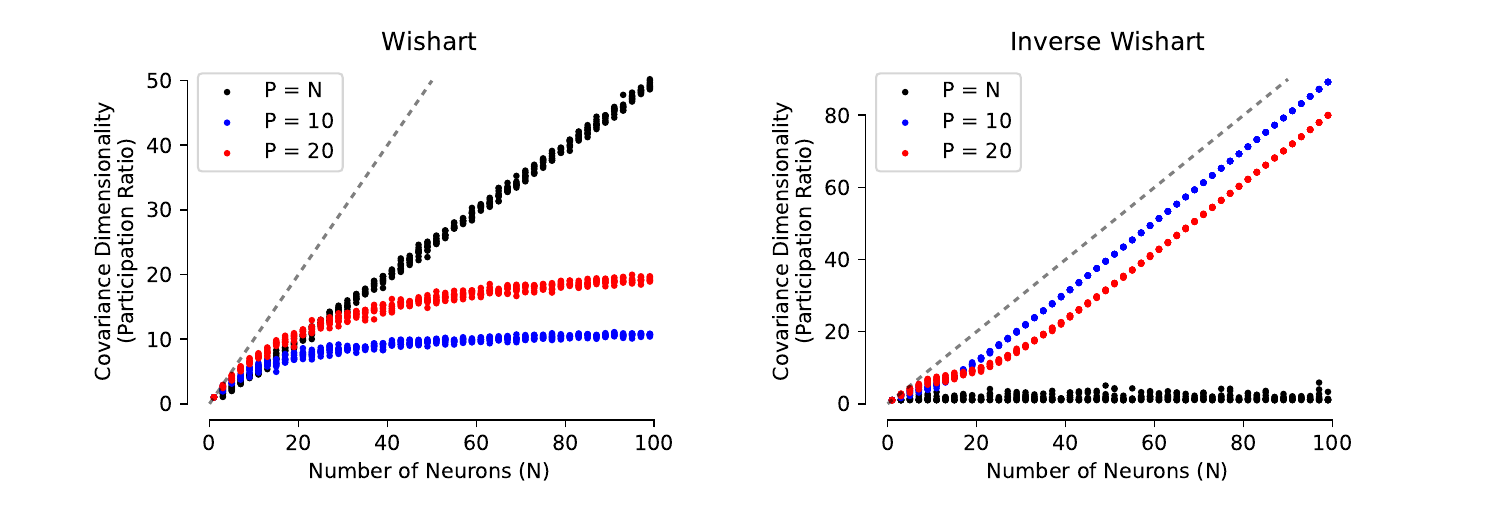}
\caption{
The dimensionality of covariance matrices sampled from the prior distribution under Wishart and Inverse Wishart models (black) as a function of the number of neurons ($N$).
The dimensionality of the covariance matrix is quantified by the participation ratio: $\sum_i (\lambda_i)^2 / \sum_i \lambda_i^2$, where $\lambda_i$ are the eigenvalues of the matrix.
\textit{Left}, Wishart distribution prior, $\mbSigma = \mbU \mbU + \sigma^2 \mbI$ where the elements of $\mbU \in \reals^{N \times P}$ are standard normal variables $\cN(0, 1)$.
For the full rank model ($P = N$ in black), $\sigma^2=0$.
For low-rank models ($P=10$ and $P=20$ in blue and red), $\sigma^2=0.1$. Dashed black line is the identity line.
Results for 10 random samples from the prior for each neural population size (different values of $N$).
\textit{Right}, same as \textit{left} but with Inverse Wishart distribution prior, $\mbSigma = \left ( \mbU \mbU + \sigma^2 \mbI \right )^{-1}$ where $\mbU$ is a random Gaussian matrix as described above.
}
\label{suppfig:low-rank-supplement}
\end{figure}

\begin{figure}[h]
\includegraphics[width=\linewidth]{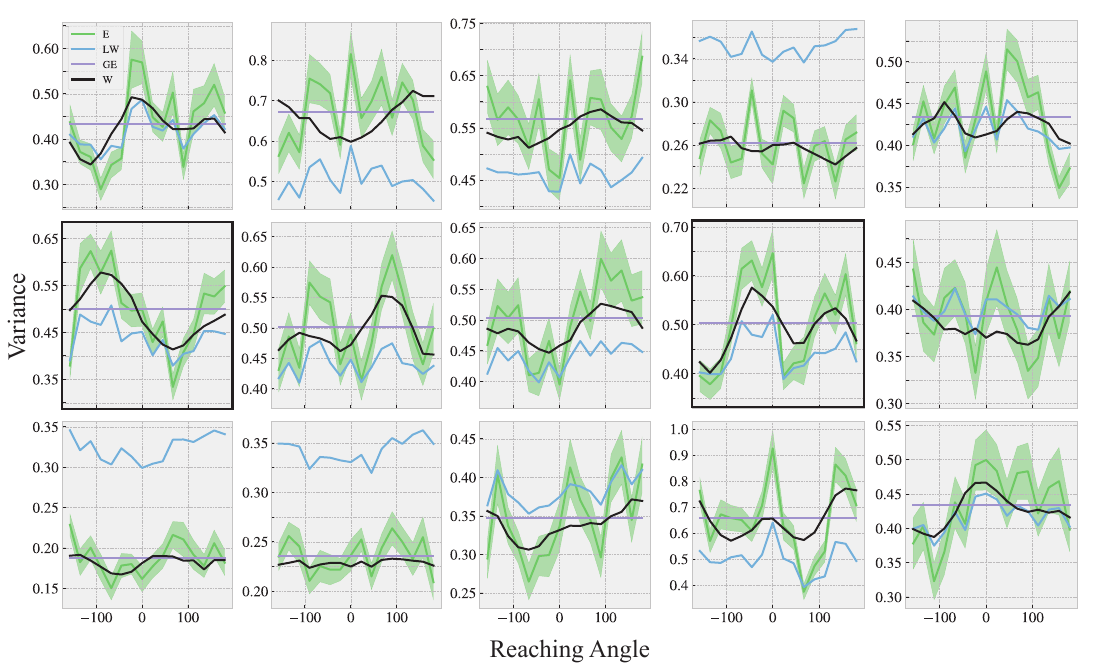}
\caption{Marginal variance estimations of 12 example neurons across reaching angles. Wishart (\emph{black}), Ledoit-wolf (\emph{blue}), grand-empirical (\emph{purple}), and empirical (\emph{green}) marginal variance estimates.  Graphical LASSO uses empirical variance to estimate the diagonal and is therefore equivalent to the empirical estimate (\emph{green line}) in this depiction  ~\cite{Friedman2007}.  }
\end{figure}

\begin{figure}[h]
\includegraphics[width=\linewidth]{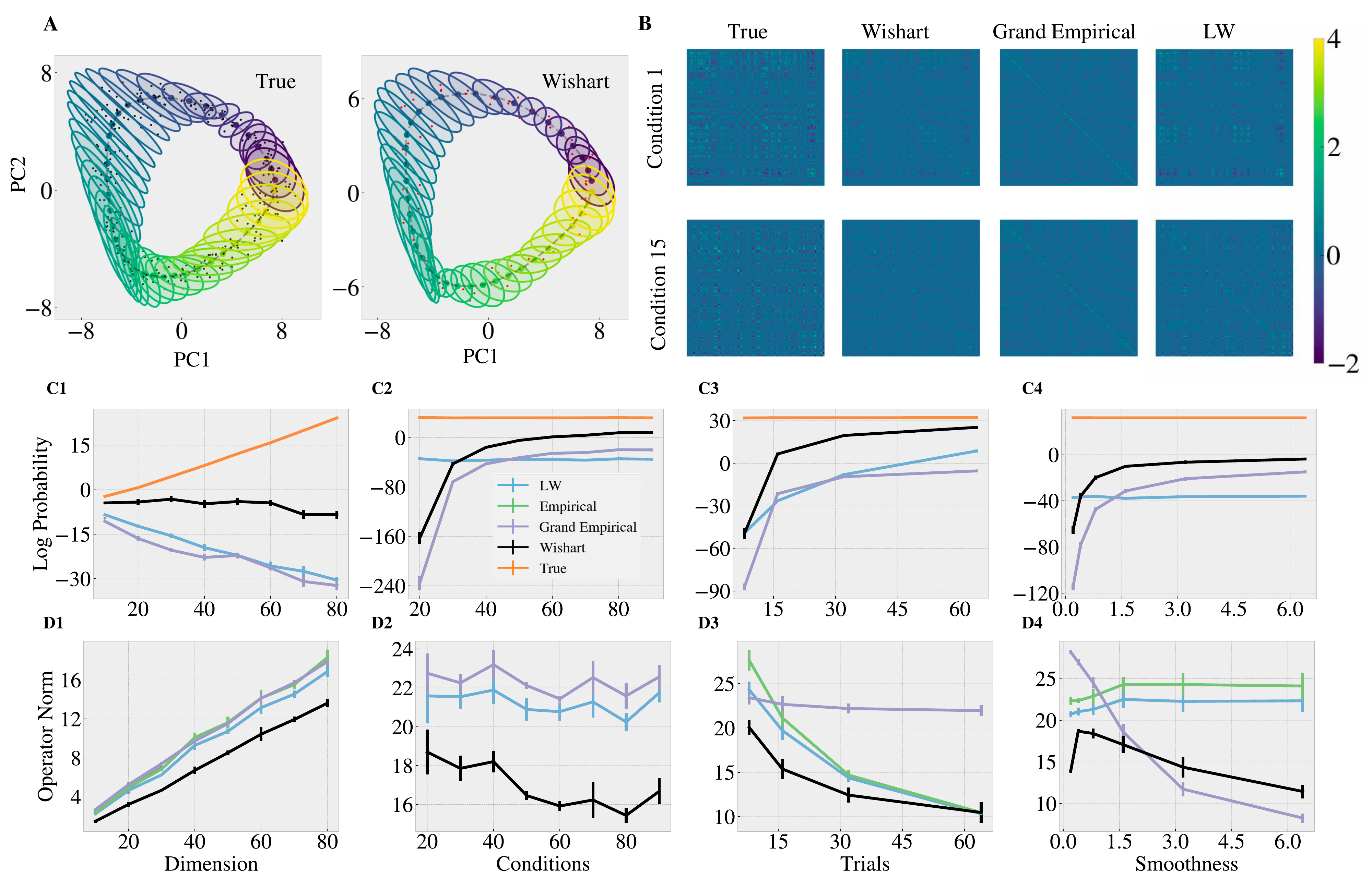}
\caption{
Results on synthetic data for $\mbL\mbL^T=\mbI$. 
(A) Similar setting to Fig.~\ref{fig:synthetic-data}, with parameters $\{N,C,K,\lambda_{\Sigma},P\}=\{100,40,10,1.0,4\}$.
Means $\mbmu(\mbx_{c})$ and covariances $\mbSigma(\mbx_{c})$ are shown in dashed lines and colored ellipses respectively for the ground truth (\emph{left}) vs. Wishart (\emph{right}) plotted in the PC space. Train and test trials $\mby_{ck}$ are shown in small black and red dots.
(B) Covariances inferred by different methods for two example conditions ($\mbx_1=0^{\circ},\mbx_{15}=168^{\circ}$). Wishart captures the low-rank structure observed in the true matrices using a small number of trials per condition.
(C1-C4) Log probability on held-out trials as a function of the number of dimensions (10-90), conditions (10-90), trials (8,16,32,64), and covariance smoothness (0.2, 0.4, 0.8, 1.6, 3.2, 6.4). For each plot, we vary one parameter and fix other parameters to $\{N,C,K,\lambda_{\Sigma},P\}=\{100,40,10,1.0,4\}$. Error bars represent median $\pm$ SEM across 5 independent runs for each parameter configuration.
(D1-D4) Mean operator norm between the true and inferred covariances showing agreement with the log probability results.
}
\label{suppfig:synthetic:eye}
\end{figure}

% \subsection{Validation of the Number of Components}

\begin{figure}[h]
\includegraphics[width=1\linewidth]{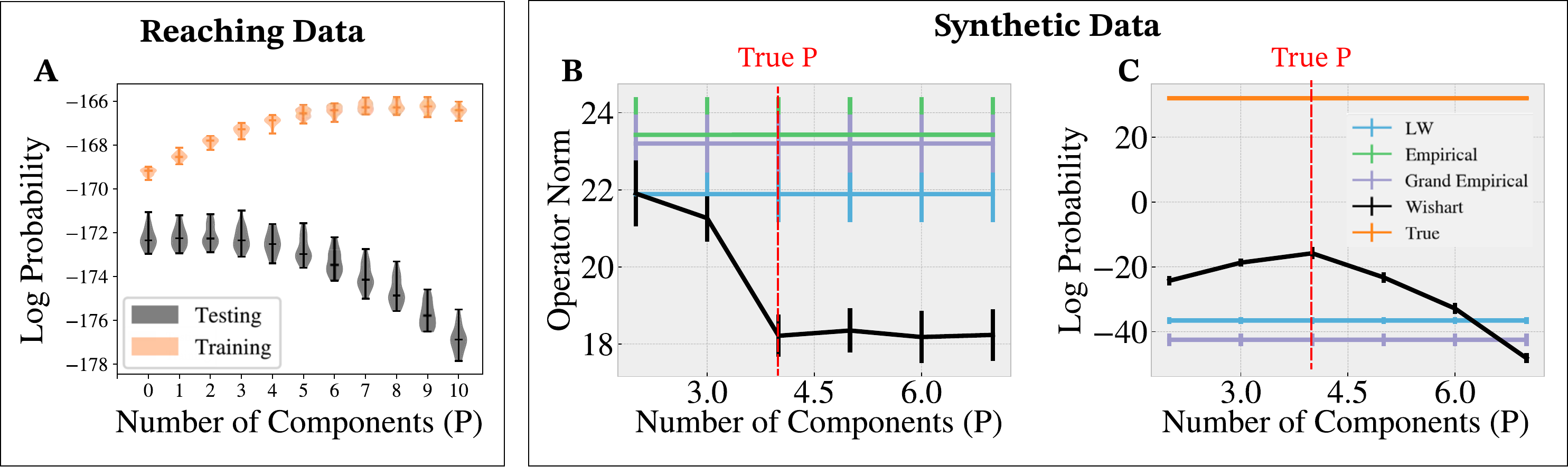}
\caption{
Validation of the number of components ($P$).
(A) Log probabilities of training and hold-out trials on 3-ring stimulus arrangement of the primate reaching task, across a range of components.  Probability on the held-out set begins to decline after $P=2$, while the probability of the training set continues to increase, evidence of overfitting to the training set at higher $P$.  
(B-C) Results on synthetic data generated from ground truth model with $P_{\text{true}}=4$; Wishart model fitted with varying $P$ (shown on the x-axis) achieving the lowest operator norm error (B) and the highest likelihood (C) when $P$ coincides $P_{\text{true}}$.
}
\label{suppfig:validation-p}
\end{figure}

\begin{figure}[h]
\includegraphics[width=\linewidth]{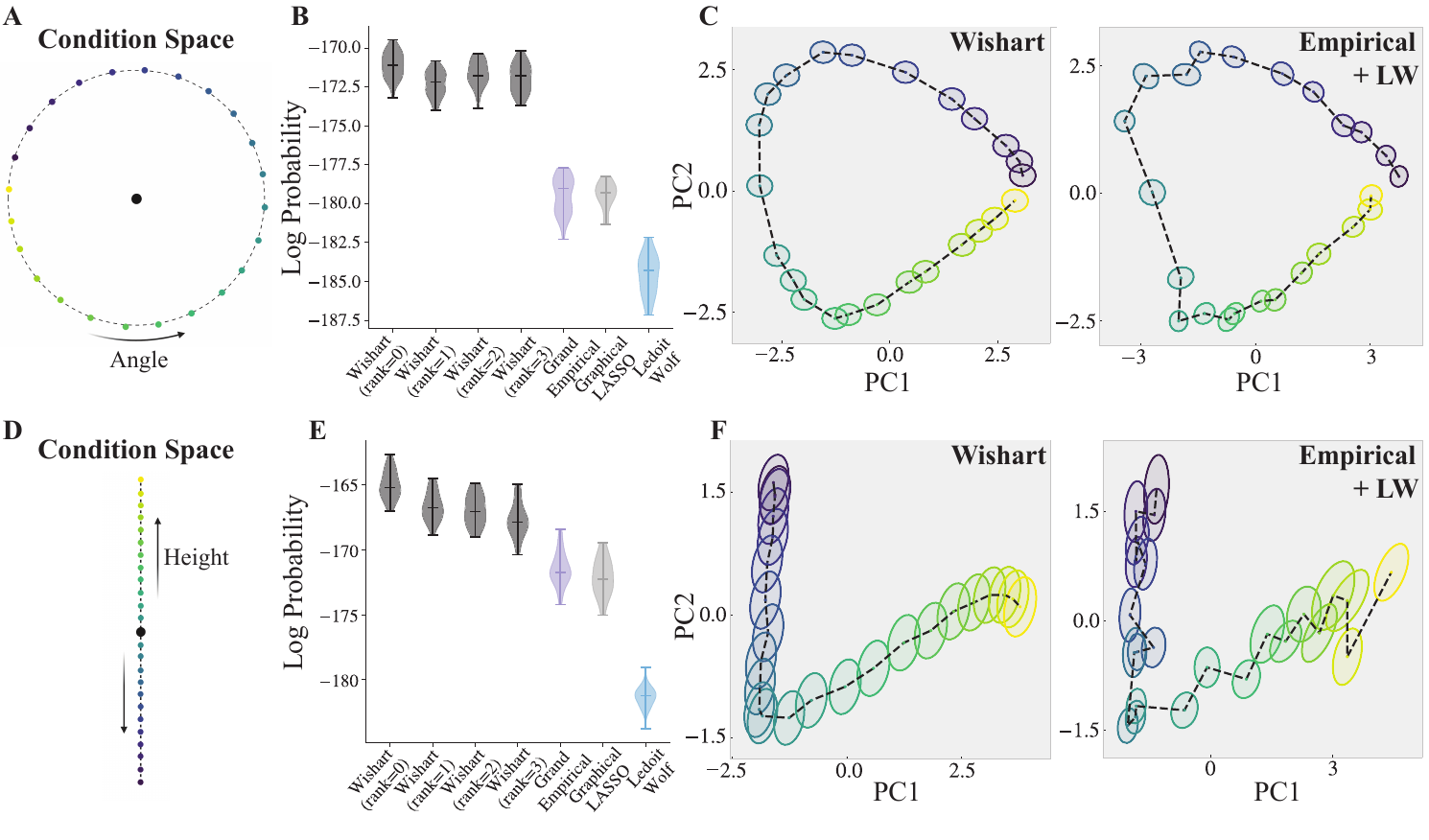}
\caption{Results on additional target configurations of primate reaching task. 
(A-C) Reaching targets are arranged on a single ring with 24 targets. This dataset contains 60 trials per reaching condition - 80\% were used as training trials, and the remaining 20\% for testing.  
A $2\pi$-periodic kernel 
% (eq.~\ref{eq:kernel-periodic}) 
was used to model smoothness around the ring.  
Scale parameters were identical to those used in Fig. 3, as was the angular smoothness parameter. 
(A) Reaching targets are arranged equidistantly around a ring with a radius of 8cm.   
(B) Log probability of hold-out trials, over 30 random folds.
(C) Wishart (\textit{left}) and Ledoit-Wolf (\textit{right}) estimates of mean and covariance projected onto the first two principal components.  
(D-F) Reaching targets are arranged on a vertical line with 24 targets.
The dataset contains 70 total trials per condition, which are similarly divided into 80\% training and 20\% testing trials.  
We used  a squared exponential kernel 
% (eq.~\ref{eq:kernel-squared-exp}) 
to model smoothness along the line.  
Scale parameters were again identical to the ones in Fig. 3.  
(D) Reaching targets are equally spaced 1cm apart along the vertical line.  
(E,F) are the same as (B,C) for the linear arrangement.
Our results are in line with \cite{Even-Chen2019-fi} Figure 2A.}
\end{figure}

\begin{figure}[h]
\includegraphics[width=\linewidth]{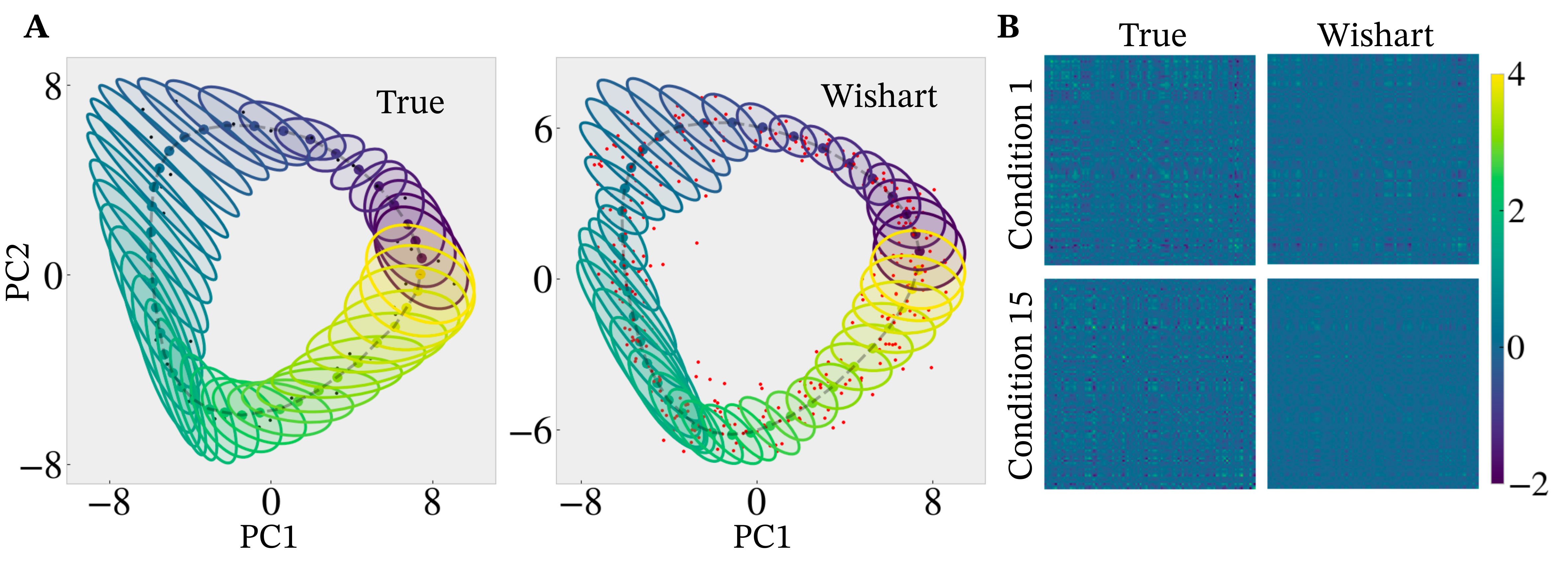}
\caption{
Inferring covariances using a single trial per condition.
(A) Left: Synthetic data generated from the ground truth model, with a single training trial per condition (black dots). Right: Covariances inferred by the Wishart model overlaid on the test trials (red dots).
(B) True and inferred covariances for two example conditions (1, 15). Wishart recovers the coarse structure of the true covariances using a single trial per condition.
}
\end{figure}

\clearpage

% \begin{figure}[h]
% \includegraphics[width=\linewidth]{figs/supp_overfitting.pdf}
% \caption{}
% \label{suppfig:overfitting-supplement}
% \end{figure}

\section{Supplemental Methods}

\subsection{Simulation Details}
\label{supp:simulation_details}

Data used in Fig.~\ref{fig:synthetic-data}A1-A5 was simulated according to the generative model in equation~\ref{eq:gaussian-wishart-priors} in 100 dimensions. The parameters used in this simulation are as follows: 100 neurons ($N=100$), 40 conditions ($C=40$), 10 trials per condition split into 8 training and 2 testing trials per condition ($K=8$). The mean covariance matrix $L$ was generated by exponentially decaying eigenvalues of rank 2 to mimic the low-dimensionality of neural responses. The condition space was parameterized by a periodic value ranging from 0-360 (akin to reach angle or visual grating orientation) and hence periodic kernel was used for this analysis. We set the number of components to 2 ($P=2$) for both the generative and inference models. The diagonal scaling of the kernel is set to 0.001 ($\gamma_{\bs{\mu}},\gamma_{\bs{\Sigma}}=0.001$) for both mean GP and covariance WP kernels. We added a diagonal matrix of 0.1 times identity to all per-condition covariance matrices to ensure they are PSD. The smoothness parameter for both GP and WP was set to 1 ($\lambda_{\bs{\mu}},\lambda_{\bs{\Sigma}}=1$). We used multivariate Gaussian conditional likelihood and drew IID samples to generate train and test trials for each condition. We used GP mean estimation in~\ref{fig:synthetic-data}A2 and empirical mean estimation for~\ref{fig:synthetic-data}A3-A5.

We performed variational inference using mean field normal distribution as the approximating family. The number of particles for importance weighting was set to 1. We performed SGD optimization using \texttt{Adam} optimizer with the step size of 0.001 using 50,000 iterations without mini-batching. In all experiments, we initialized the mean of the variational family according to the per-condition empirical means of the data and initialized the inference scale matrix according to the Cholesky factorization of the grand-empirical covariance.

For other panels in Fig.~\ref{fig:synthetic-data} we used a similar scheme to the one described above. However, in every panel, we changed one parameter of the generative model and kept all the other parameters constant. We varied the number of dimensions between 10 and 90 with steps of 10, number of conditions between 20 and 100 with steps of 10, number of trials per condition in $\{8, 16, 32, 64\}$ and the smoothness parameter for the covariance in $\{.2, .4, .8. 1.6, 3.2, 6.4\}$. The fixed parameters in each plot were chosen as $N=100, C=30, N=10,\lambda_{\bs{\Sigma}}=1$. In all the log probability plots (Fig.~\ref{fig:synthetic-data}B) we used empirical mean estimation to compare the performance of covariance estimators in isolation.
All the results in Fig.~\ref{fig:synthetic-data} are based on the held-out test trials.

\subsection{Inverse Wishart Process}
\label{supp:inverse_wishart}

The inverse Wishart process is defined as follows:
\begin{equation}
\label{eq:inverse-wishart-with-diag-component}
\mbSigma(\mbx) = \left ( \mbL \left ( \mbU(\mbx) \mbU(\mbx)^\top + \mbLambda(\mbx) \right ) \mbL^\top \right)^{-1}.
\end{equation}
Since the mapping $\mbSigma \mapsto \mbSigma^{-1}$ is continuous over (strictly) positive definite matrices, imposing smoothness on $\mbSigma^{-1}(\mbx)$ implies some degree of smoothness in $\mbSigma(\mbx)$.
Nonetheless, the Wishart and inverse-Wishart process specify distinct prior distributions, and it is an empirical question of which one is better suited to any specific circumstance (see Fig.~\ref{suppfig:low-rank-supplement} for the dimensionality of covariance matrices sampled from the priors of Wishart and inverse-Wishart models).
Importantly, one can perform inference in eq.~(\ref{eq:inverse-wishart-with-diag-component}) without representing and inverting the covariance matrix explicitly, which would lead to numerical instabilities.
Therefore, eq.~(\ref{eq:inverse-wishart-with-diag-component}) may be preferable when the inverse covariance is of greater interest than the covariance---e.g., in calculations of linear Fisher information~\cite{Moreno2014}.

\subsection{Inference}
\label{supp:inference}
The inference algorithm that we propose is similar to ~\cite{ober2021variational}, except that we do not utilize inducing points as the number of conditions in neuroscience experiments is often small. Instead, we directly perform variational inference on the latent variables $\{ \mbU(\mbx_c) \}_{c=1}^C$ and $\{ \mbmu(\mbx_c) \}_{c=1}^C$.
This allows us to easily switch between MAP estimates of the variables and posterior inference by choosing delta or normal distribution as approximating variational family with the latter introducing further regularization.

We denote the generative parameters by $\bs{\theta}$, which in our case only contains the $\mbL$ matrix.
Given the choice of $q_{\bs{\phi}}$, we would like to find 
$q_{\bs{\phi}}\left (
\{ \mbU(\mbx_c),
\mbmu(\mbx_c) \}_{c=1}^C
\right )$
to best approximate
$p_{\bs{\theta}} \left (
\{ \mbU(\mbx_c),
\mbmu(\mbx_c) \}_{c=1}^C
\mid \{\mby_{ck}\}_{c=1,k=1}^{C, K}
\right )$
while optimizing over the parameters $\bs{\theta}$.
We follow the literature on stochastic variational inference and derive the ELBO cost function for this problem.
\begin{gather*}
    \mathcal{L}(\bs{\phi},\bs{\theta}) = \E_{q}\big[\log p_{\bs{\theta}} \left (
\{ \mbU(\mbx_c),
\mbmu(\mbx_c) \}_{c=1}^C
, \{\mby_{ck}\}_{c=1,k=1}^{C, K}
\right )
- \log q_{\bs{\phi}}\left (
\{ \mbU(\mbx_c),
\mbmu(\mbx_c) \}_{c=1}^C
\right ) \big]
\end{gather*}
Using the reparameterization trick, we achieve an empirical estimate of the gradient of loss using samples from an independent noise.
\begin{gather*}
\bs{z}\coloneqq \left ( \{ \mbU(\mbx_c) \}_{c=1}^C,\{ \mbmu(\mbx_c) \}_{c=1}^C \right) \quad \bs{z} = h_{\bs{\phi}}(\bs{\epsilon}) \\
    \nabla_{\bs{\phi},\bs{\theta}} \mathcal{L}(\bs{\phi},\bs{\theta}) = \E_{\bs{\epsilon}}\big[\nabla_{\bs{\phi},\bs{\theta}} \log p_{\bs{\theta}} \left (
    \{\mbx_{c}\}_{c=1}^{C},\{\mby_{ck}\}_{c=1,k=1}^{C, K}
    ,h_{\bs{\phi}}(\bs{\epsilon})
    \right)
    - \nabla_{\bs{\phi}} \log q_{\bs{\epsilon}}(h_{\bs{\phi}}(\bs{\epsilon})) \big]
\end{gather*}
In practice, we use Monte Carlo estimates of the expectation above by sampling from an independent noise distribution and evaluating the gradient inside the expectation for those samples. The number of samples used for estimating the gradient introduces a trade-off between the accuracy of the estimation computation time. We empirically find that a single sample is often enough to achieve accurate estimates. We choose \texttt{Adam} optimizer with the step size $0.001$ to perform our stochastic gradient-based optimization.

\subsubsection{Computational Complexity}
In order to examine the computational complexity of our variational framework, we first expand the gradient calculation. Notice that the complexity of our model is not the same as out-of-the-box stochastic variational models, due to the specific structure in our prior model, namely Gaussian and Wishart Processes.
\small
\begin{align*}
\nabla_{\bs{\phi},\bs{\theta}} \mathcal{L}(\bs{\phi},\bs{\theta}) &= \E_{\bs{\epsilon}}\big[\nabla_{\bs{\phi},\bs{\theta}} \log p_{\bs{\theta}}\left (
\{\mbx_{c}\}_{c=1}^{C},\{\mby_{ck}\}_{c=1,k=1}^{C, K}
,h_{\bs{\phi}}(\bs{\epsilon})
\right ) - \nabla_{\bs{\phi}} \log q_{\bs{\epsilon}}(h_{\bs{\phi}}(\bs{\epsilon})) \big] \\ 
&\approx \sum_{c,k} \nabla_{\bs{\phi},\bs{\theta}}  \log p_{\bs{\theta}}\left(
\mbx_{c},\mby_{ck}
\mid h_{\bs{\phi}}(\bs{\tilde{\epsilon}})
\right) +
\nabla_{\bs{\phi},\bs{\theta}} \log p_{\bs{\theta}}(h_{\bs{\phi}}(\bs{\tilde{\epsilon}}))  - \nabla_{\bs{\phi}} \log \N(h_{\bs{\phi}}(\bs{\tilde{\epsilon}});\bs{0},\mathbb{1})
\end{align*}
\normalsize
The first term (log observation), requires inverting and the determinant of the per-condition covariance matrices, which has a complexity of $\mathcal{O}(CN^3)$ and $K$ matrix multiplications with the covariance matrices which has a complexity of  $\mathcal{O}(CKN^2)$. The second term (log prior) requires inverting $P$ matrices that are of size $C \times C$ and, therefore have the complexity of $\mathcal{O}(C^3P)$. Notice that if the kernel parameters are not optimized, the kernel matrix only needs to be evaluated once, not for every iteration. 
% Finally, this evaluation of the gradient is for a single sample $\bs{\tilde{\epsilon}}$ from the variational family. 
% In practice, we use importance weighting and use a single sample.
% which adds a multiplicative $K$ to the computational complexity.

Overall, the complexity is given by $\mathcal{O}(CN^3+CKN^2+C^3P)$, which can be improved by mini-batching (turning the multiplicative term $CK$ into the batch size). Furthermore, if we assume that we have few conditions $C \sim \mathcal{O}(1)$ and the number of neurons scales quadratically with the number of trials $K \sim \mathcal{O}(\sqrt{N})$ and use mini-batching this reduces the overall complexity to $\mathcal{O}(N^3)$. Notice that these computations are required for a single iteration of SVI.

We observed that on GPU it takes about 80 seconds to fit a dataset with 100 neurons, 80 conditions, and 32 trials per condition. Notice that we run 10000 iterations of our optimization algorithm, therefore the run time for such a dataset is about 8 milliseconds per iteration. The runtime for other algorithms for a similarly sized dataset are the following (in seconds): PoSCE: 45, Graphical Lasso: 24, Ledoit-Wolf: 0.3, and Empirical: 0.3. All other algorithms were run on CPU as GPU implementations are not available. 

Notice that our model has fundamentally new capabilities that are not present in simple baselines. In particular, we can infer a continuous manifold for the mean and covariance of neural responses (see Fisher Information analysis), generalize to entirely unseen conditions, and quantify our uncertainty in a Bayesian framework. A fair comparison between inference times should take this difference into account.

\subsection{Posterior Predictive Distribution}
Posterior predictive distributions provide a principled approach in our model to sample means and covariances in training and unseen test conditions.
Posterior predictive defines a distribution over unseen test pair $(\bs{x}^*,\bs{y}^*)$ conditioned on the training data with the inferred posterior marginalized out. In practice, posterior predictive likelihoods can be estimated using samples from the approximate posterior distribution in the following way.  

\begin{align*}
    p_{\bs{\hat{\theta}}}(\bs{x}^*,\bs{y}^* | \mathcal{D}) &= \int p_{\bs{\hat{\theta}}} (\bs{x}^*,\bs{y}^*|\bs{z}) p_{\bs{\hat{\theta}}}(\bs{z}|\mathcal{D}) d\bs{z} =
    \E_{p_{\bs{\hat{\theta}}}(\bs{z}|\mathcal{D})}  [ p_{\bs{\hat{\theta}}}(\bs{x}^*,\bs{y}^*|\bs{z})] 
    \\ &\approx \frac{1}{S} \sum_{\bs{z}_s \sim p_{\bs{\hat{\theta}}}(\bs{z}|\mathcal{D})}  p_{\bs{\hat{\theta}}}(\bs{x}^*,\bs{y}^*|\bs{z}_s) 
    \approx \frac{1}{S} \sum_{\bs{z}_s \sim q_{\bs{\hat{\phi}}}(\bs{z})} p_{\bs{\hat{\theta}}}(\bs{x}^*,\bs{y}^*|\bs{z}_s) 
\end{align*}
where every $\bs{z}_s$ for $s=\{1,\dots,S\}$ are $S$ samples from the inferred posterior on 
$\left ( \{ \mbU(\mbx_c) \}_{c=1}^C,\{ \mbmu(\mbx_c) \}_{c=1}^C \right)
$
and $p_{\bs{\hat{\theta}}}(\bs{x}^*,\bs{y}^*|\bs{z}_s)$ is the model likelihood conditioned on the sampled $\bs{z}_s$. If $\bs{x}^*$ is among the training conditions, this likelihood is a straightforward evaluation of a multivariate Gaussian with parameters given by
$\mbU(\mbx^*=\mbx_c),\mbmu(\mbx^*=\mbx_c)$.
If the test condition is not observed among the training conditions, we can apply the Bayes rule again and estimate its likelihood. Notice that this interpolation property is allowed in our model due to the Gaussian and Wishart process priors. Sampling from the model and evaluating its likelihood for unobserved conditions is not possible in previous models since the models render different conditions independent and do not consider a smoothness prior.
\begin{align*}
p\left (\bs{x}^*,\bs{y}^*
\mid
 \{ \mbU(\mbx_c), \mbmu(\mbx_c) \}_{c=1}^C 
\right) = \E_{p \big(
\mbU(\mbx^*),\mbmu(\mbx^*)
\mid 
\{ \mbU(\mbx_c), \mbmu(\mbx_c)  \}_{c=1}^C 
\big)} [p(\bs{y}^* \mid
\mbU(\mbx^*),\mbmu(\mbx^*)
)]
\end{align*}
where samples from $p \big(
\mbU(\mbx^*),\mbmu(\mbx^*)
\mid 
\{ \mbU(\mbx_c), \mbmu(\mbx_c)  \}_{c=1}^C 
\big)$ are drawn from a multivariate Gaussian with a covariance provided by the kernel.

\subsection{Spiking Count Model}
\label{supp:poisson}
It has been argued that for a dataset of spike counts a gain-modulated multivariate Poisson likelihood model better represents first and second-order statistics of the data while providing flexibility for incorporating and estimating neural correlations~\cite{goris2014partitioning}. 
\begin{gather*}
\label{eq:poisson}
    \bs{\mu}(\bs{x}) \sim \mathcal{GP}^{N} \quad \bs{\Sigma}(\bs{x}) \sim \mathcal{WP}^{lrd}(\bs{L},P) \\
    \bs{g}_{ck}|\bs{x}_c \sim \N(\bs{g}|\bs{\mu}_c,\bs{\Sigma}_c)\quad
    \bs{y}_{ck}|\bs{g}_{ck} \sim \text{Poiss}(\bs{y}|\texttt{softplus}(\bs{r}+\bs{g}_{ck}))
\end{gather*}
where $\bs{g}_{ck}$ is a vector of gains sampled from a multivariate normal distribution, $\bs{y}_{ck}$ is the vector of spike counts for condition $c$, and trial $k$, and $\bs{r}$ is the vector of condition-independent baseline spike rates for individual neurons that we optimize. The graphical model of this statistical model is shown in Fig. 1F.
\paragraph{Inference Details} We highlight the differences between running inference in the Normal and Poisson models. The \textbf{latent variables} of the Poisson model include $\bs{g}_{ck}$ and our mean field distribution factorizes over $\bs{\mu}_{1:C},\bs{\Sigma}_{1:C},\bs{g}_{1:C,1:K}$. Hence, our posterior is defined over the joint space of these variables. The \textbf{parameters} of the generative model in addition to the prior parameters (i.e. $\bs{L}$) include the likelihood rate parameters as well $\bs{\theta}= \{\bs{L},\bs{r}\}$. As before, we optimize the ELBO cost function and run the stochastic variational inference with $\texttt{Adam}$ optimizer. We observed that a larger \textbf{step size} (0.005) and more \textbf{optimization iterations} (50000) are often needed to achieve the best results.

\paragraph{Summary Statistics} In the Poisson model the means and covariances are defined and estimated in the gain space instead of the spike count space. However, our statistical model allows for sampling the means and covariances in the spike count space using a simple Monte Carlo scheme.
\begin{equation*}
    \E ss(\bs{y}) \approx \frac{1}{S} \sum_{\substack{
    \bs{\mu}_s,\bs{\Sigma}_s \sim q_{\hat{\bs{\phi}}}(\bs{\mu},\bs{\Sigma}|\mathcal{D})\\
    \bs{g}_s \sim p(\bs{g}|\bs{\mu}_s,\bs{\Sigma}_s)\\
    \bs{y}_s \sim p_{\hat{\bs{\theta}}}(\bs{y}|\bs{g}_s)
    }} ss(\bs{y}_s)
\end{equation*}
for any summary statistic function $ss$. If we choose $ss(\bs{y})=\bs{y}$, the equation above will return the estimate of mean spike counts. If we choose $ss(\bs{y}) = (\bs{y}-\E\bs{y})(\bs{y}-\E\bs{y})^T$ it will return the estimate of the covariance of the spike counts. 

\begin{figure}[h]
\centering
\includegraphics[width=.6\linewidth]{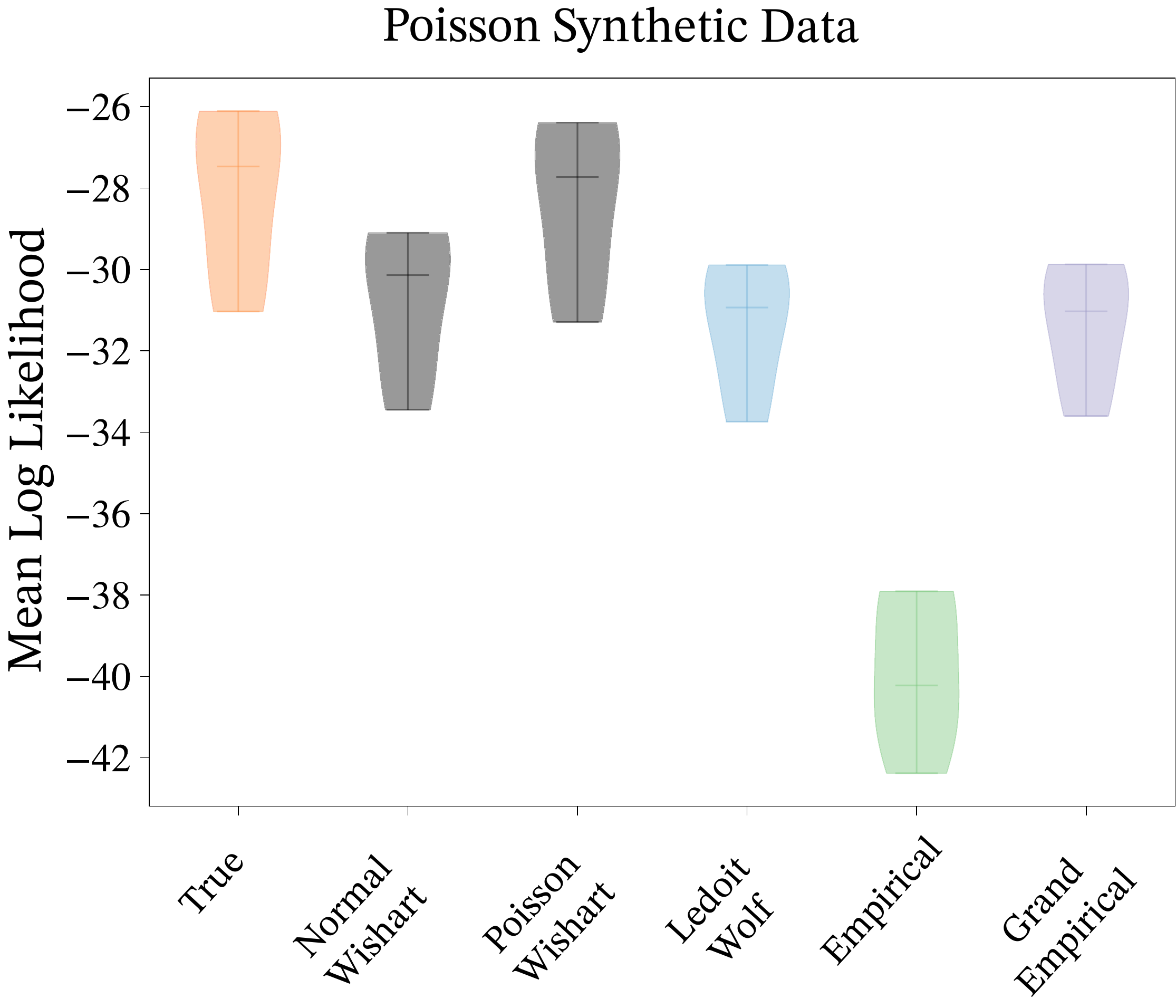}
\caption{
Results from synthetically generated Poisson firing rates. We generated data from the model in eq.~\ref{eq:poisson} with 20 neurons, 30 conditions, and 50 trials per condition. We then inferred means and covariances separately using the Normal and Poisson models and evaluated the test log-likelihoods of the held-out trials. The Poisson model produced log-likelihood values closest to the ground truth and outperformed the Normal model as well as other compared methods.
}
\label{suppfig:poisson}
\end{figure}

\paragraph{Likelihood Evaluation}
To compare which of the Poisson or Normal models is a better fit to the data, we need to compare the corresponding likelihoods. However, not having access to the gain parameter in the Poisson model makes the exact likelihood evaluation intractable. Instead, we will marginalize out the gain parameter using the following integral similar to the posterior predictive.
\begin{align*}
    p(\bs{y}|\mathcal{D}) &= \int p_{\hat{\bs{\theta}}}(\bs{y}|\bs{g})p(\bs{g}|\bs{\mu},\bs{\Sigma}) p_{\hat{\bs{\theta}}}(\bs{\mu},\bs{\Sigma}|\mathcal{D}) d\bs{g}d\bs{\mu}d\bs{\Sigma} \\
    & \approx \int p_{\hat{\bs{\theta}}}(\bs{y}|\bs{g})p(\bs{g}|\bs{\mu},\bs{\Sigma}) q_{\hat{\bs{\phi}}}(\bs{\mu},\bs{\Sigma}|\mathcal{D}) d\bs{g}d\bs{\mu}d\bs{\Sigma}
\end{align*}
which in practice is performed using Monte Carlo samples from the corresponding distributions. A comparison of the log-likelihood values achieved by the Poisson and Normal models for Poisson-generated data is shown in Fig.~\ref{suppfig:poisson}.

\subsection{Fisher Information Estimation}
\label{supp:fi}
Suppose $f:\R^M \rightarrow \R$ is distributed according to the Gaussian Process $\mathcal{GP}(f;\mu,K)$ where $\mu,K$ are mean and kernel functions, and assuming that $\frac{\partial \mu}{\partial x_i}(\bs{x})$ and $\frac{\partial^2 K}{\partial x_i \partial x'_i}(\bs{x},\bs{x})$ exist for all $i$ and any $\bs{x},\bs{x'}$ then for $f$ and its gradient $f'$ we have the following result thanks to the linearity of differentiation operation.
\begin{equation*}
    p(f,\nabla f)  = \mathcal{GP}\bigg(
    \begin{bmatrix}
        f\\
        f'
    \end{bmatrix}; 
    \begin{bmatrix}
        \mu\\
        \mu'
    \end{bmatrix},
    \begin{bmatrix}
        K &  \frac{\partial K}{\partial \bs{x}}(\bs{x'},\bs{x})^T\\ \frac{\partial K}{\partial \bs{x}}(\bs{x},\bs{x'}) & \frac{\partial^2 K}{\partial \bs{x} \partial \bs{x'}}(\bs{x},\bs{x'})
    \end{bmatrix}
    \bigg)
\end{equation*}

Since the posterior GP is another GP with data-dependent mean and kernel functions, we can jointly sample a function and its gradient or compute the gradient of the posterior mode.

Given this result, we can sample from the posterior derivative and compute the uncertainty associated with it. In addition, we can compute derived quantities such as Fisher Information as described in the main text.

\end{appendices}

\end{document}